\documentclass[%
 reprint,
superscriptaddress,
 amsmath,amssymb,
aps,
pra,
]{revtex4-2}

\usepackage{graphicx}
\usepackage{dcolumn}
\usepackage{subcaption}
\usepackage{caption}
\usepackage{ragged2e}
\usepackage{bm}% bold math
\usepackage{tensor}
\usepackage{physics}
\usepackage{booktabs}
\usepackage{rotating}
\usepackage{makecell}
\usepackage{array}

\usepackage{cleveref}

\usepackage[most]{tcolorbox} % Required for theorem boxes

\newtcbtheorem[number within=section]{mytheorem}{Theorem}%
{colback=blue!5,colframe=blue!35!black,fonttitle=\bfseries}{th}
\usepackage{amsthm}

\makeatletter
\def\bstctlcite{\@ifnextchar[{\@bstctlcite}{\@bstctlcite[]}}
\def\@bstctlcite[#1]#2{\@bsphack
  \@for\@citeb:=#2\do{%
    \edef\@citeb{\expandafter\@firstofone\@citeb}%
    \if@filesw\immediate\write\@auxout{\string\citation{\@citeb}}\fi}%
  \@esphack}
\makeatother
\newcommand{\levicivita}{}% initialize
\def\levicivita#1#{\tensor#1{\epsilon}}

\begin{document}

\bstctlcite{APS:BSTcontrol}

\preprint{APS/123-QED}

\title{The Initiation of Superradiance from Different Collective-Spin States}

\author{Adnan N. Alabbar}
\thanks{Corresponding author: Adnan.alabbar94@tamu.edu}
\affiliation{Institute for Quantum Science and Engineering, Texas A\&M University, College Station, Texas 77843, USA}
\affiliation{Department of Physics and Astronomy, Texas A\&M University, College Station, Texas 77843, USA}

\author{Zhenghao Zhang}
\affiliation{Institute for Quantum Science and Engineering, Texas A\&M University, College Station, Texas 77843, USA}
\affiliation{Department of Physics and Astronomy, Texas A\&M University, College Station, Texas 77843, USA}

\author{Girish S. Agarwal}
\affiliation{Institute for Quantum Science and Engineering, Texas A\&M University, College Station, Texas 77843, USA}
\affiliation{Department of Physics and Astronomy, Texas A\&M University, College Station, Texas 77843, USA}
\affiliation{Department of Biological and Agricultural Engineering, Texas A\&M University, College Station, Texas 77843, USA}

\date{\today}% It is always \today, today,

\begin{abstract}
\noindent Superradiance is an extensive cooperative spontaneous emission phenomenon, exhibited by some atomic collective-spin states.
However, distinct initial states differ in their decay dynamics.
Dicke states $|j,m\rangle$ with distinct numbers of excitations $n=m+j$, driven by vacuum fluctuations, have their peak emission intensity shifted in time.
Rotating Dicke states relative to the decay axis introduces an interesting parity structure that affects the pulse profile and photon correlations. 
Squeezed-bath prepared states undergo a squeezing-controlled crossover to the rotated Dicke states, making the emission character dependent on the amount of squeezing transferred from light to the atomic state.
For semiclassical states with a macroscopic dipole moment, like the atomic coherent state, the emission intensity depends on their polarization.
We present detailed results on the superradiant dynamics of a representative selection of states expanded in Dicke states to highlight the initial state as an independent dynamical control parameter.
For large-$N$, we are able to predict fairly accurately the pulse profile in each case using the mean-field approximation, an approach based on the Fokker--Planck equation.
We also present comparative results on the intensity correlation function, quantify the coherent and incoherent contributions of the emission, and contrast between small and large ensembles.

\end{abstract}

\maketitle

\section{Introduction}
\noindent
Dicke superradiance \cite{PhysRev.93.99} is one of the most widely studied topics in quantum optics \cite{Rehler1971Superradiance,Gross1982, PhysRevA.4.302,PhysRevA.4.854,PhysRevA.5.1457,PhysRevA.10.1096,PhysRevLett.36.1035, PhysRevA.12.2568, Scully2009, Bojer2022, THE1PhysRevLett.131.033605, THE3PRXQuantum.5.010344, THE4Masson2022,prazeres2026}. 
A number of recent experiments
\cite{REC3Solano2017, rec5kersten2026self, rec6PRXQuantum.5.040335, rec7Araujo2016, rec8Devoe1996, rec9Kersten2023, rec10Richter2023,Douglas2026, Mlynek2014}
in atomic arrays and clouds, trapped ions, nanophotonic systems, and artificial atoms have reported superradiant effects.
Central to Dicke's prediction was that a system of $N$ two-level atoms \cite{allen1975optical} prepared initially in a state such that half the atoms were excited would exhibit a radiation rate which is $\propto N^2$-times that for the emission intensity of a single atom $I_1$, and $\propto N$-times that of a system initially prepared in a fully-excited state. 
He also predicted a pulse-width reduction by $\propto 1/N$.
Dicke used the addition of angular momenta to construct the symmetric collective-spin states now known as Dicke states.
Excitation and relaxation proceed along the Dicke ladder, whose rungs are the eigenstates $|j,m\rangle$.
Besides the Fermi golden rule calculation \cite{Agarwal_2012}, a full dynamical theory was worked out in the seventies \cite{agarwal1974quantum}, and a master equation was derived which can yield the emergence of a superradiant pulse starting with all atoms in the excited state \cite{Agarwal_Master_Equation}.

The physical understanding of superradiance in Dicke states has emerged from the point of view of nonvanishing interatomic dipole correlations \cite{yelin2005,DosSantos2016} and from the perspective of quantum interference of different paths that lead to the detection of a photon \cite{PhysRevA.84.023805}.
More recently, there has been renewed interest in exploring whether the superradiant all-excited state is entangled.
Several authors have concluded that it is separable \cite{Rosario2025, Bassler2025, Wolfe2014}.
This question is important because cooperative enhancement of radiation \textit{per se} is not evidence of multipartite entanglement . Determining whether entanglement is present therefore distinguishes the correlations sufficient to produce the enhanced emission from quantum many-body correlations \cite{Zhang2025}. 

We investigate a question that has received much less attention: 
How do the properties of the superradiant pulse depend on the initial state in which the system is prepared?
We have the option of using several states which can have quantum entanglement, and thus we can study the quantum characteristics of the emission depending on quantum correlations in the initial state.
We use the familiar Dicke states and Atomic Coherent States (ACS) as reference cases for collective emission.
Varying the initial position along the Dicke ladder of the former tells us how the excitation number controls the onset, peak, delay, and width of the pulse.
As for the latter, ACS represents the coherent, semiclassical limit of collective emission because of its macroscopic dipole and separability.
We contrast them with two families of states derived from Dicke states.
The central Rotated Dicke State (RDS) has an even-parity distribution along the Dicke ladder, and the Squeezed Dicke State (SDS) is produced from the RDS by a squeezing-dependent exponential reweighting.
The emissions from RDS and SDS provide a means of determining how pair-correlated state preparations modify the resultant pulse and photon statistics.

The organization of the paper is as follows: 
In Sec. II, we recall some of the main features of the dynamical equation for superradiance. 
Sec. III studies Dicke states with different initial excitations; 
Sec. IV presents RDS, where the central state exhibits the property that only levels with even numbers of excitations are populated; 
Sec. V investigates SDS, a class of Dicke states that are the steady states of the squeezed bath master equation, and whose intensity profile is determined from the amount of squeezing; 
and finally, in Sec. VI, we bring to light new features of emission from the system prepared in an ACS.

The numerical solution of the master equation yields a superradiant pulse in each case. 
We develop analytical transfer coefficients for the master equation in Appendix A and a large-$N$ mean-field (MF) approximation in Appendix B by converting the rate equation into a Fokker--Planck equation \cite{risken1996fokker,AgarwalPS1969,Dimauro2025}. 
We also give results for the intensity-intensity correlations and dipole fluctuations, which quantify the incoherent part of the emitted radiation. 
These quantities are compared in the Discussion section, together with a fitting table for experimentalists.
\section{Method}
\noindent We consider a system of identical $N$ two-level atoms in a cavity with a leakage rate $2\kappa$. The atoms are on resonance with the cavity mode. The coupling, $g$, of the atom to the cavity is assumed to be much smaller than $\kappa$. In such a case, the dynamical behavior of the atoms is described in terms of the superradiance master equation \eqref{Master_Equation} for the atomic density matrix, $\partial_\tau \rho = 2\hat J_-\rho\hat J_+ -\{\rho,\hat J_+\hat J_-\}$, where $\tau=\Gamma t$ is the normalized time, and the dissipation rate is $\Gamma = g^2/\kappa$. Here, $\hat J$ is the collective angular momentum operator for $N$ two-level atoms with $j=N/2$ being the collective spin of the system \cite{allen1975optical, PhysRevA.2.883, PhysRevA.2.889}. We thus work in a totally symmetric space. We will use the basis states $|j,m\rangle$, with $\hat J^2|j,m\rangle =J^2|j,m\rangle$ (where $J^2=j(j+1)$), and $\hat J_z|j,m\rangle =m|j,m\rangle$ with $-j\leq m \leq +j$.

Though the intensity gives the mean emission rate, it does not determine the photon statistics of the emitted radiation.
Hence, we calculate the normalized second order correlation (Glauber) function $g^{(2)}$ \cite{glauber1963theory, PhysRevA.2.1607} defined in \eqref{Glauber_function_Analytical}, which compares the two-photon coincidence probability with the product of the corresponding one-photon probabilities \cite{scully1997quantum,Agarwal_2012,milonni2019introduction}.
The Glauber function separates states with similar intensity profiles, such as the CDS and the $x$-polarized ACS, and exposes the enhanced normalized pair-emission fluctuations of the RDS and SDS.
In terms of photon-counting, its numerator is the factorial moment $\langle \hat n(\hat n-1)\rangle$, where $\hat n=\hat a^\dagger \hat a$.
The subtraction excludes counting one photon as both members of the same coincidence pair.
Thus, two sources with similar intensities (or even first-order coherence) may be distinguished by their intensity-intensity correlations, which retain fluctuation and interference information absent from the mean-field.
At equal times, $g^{(2)}<1$ denotes antibunching and nonclassicality, $g^{(2)}=1$ is the Poissonian benchmark produced by coherent light, and $g^{(2)}>1$ denotes bunching.
Naturally, $g^{(2)}$ is undefined for the ground state, or when $I=0$.
Moreover, because $g^{(2)}$ is normalized by the square of the total intensity, it becomes increasingly sensitive to numerical error and less representative of appreciable radiation as the intensity approaches zero; each curve is terminated at the last time $\tau^*$ when it reaches $10\%$ on the falling side of the peak intensity.

In intensity simulations, we use QuTiP and SciPy to compare the analytical result with our MF solution through the Normalized Root Mean Square Error (NRMSE) from 
$\eta=\sqrt{\sum_\ell \Big(I^\text{QuTiP}_{\ell}-I^{\text{MF}}_{\ell}\Big)^2 \Big/
\sum_\ell \Big(I^\text{QuTiP}_{\ell}\Big)^2}$.
\section{Dicke States}
\begin{figure}
    \centering
    \includegraphics[width=0.96\linewidth]{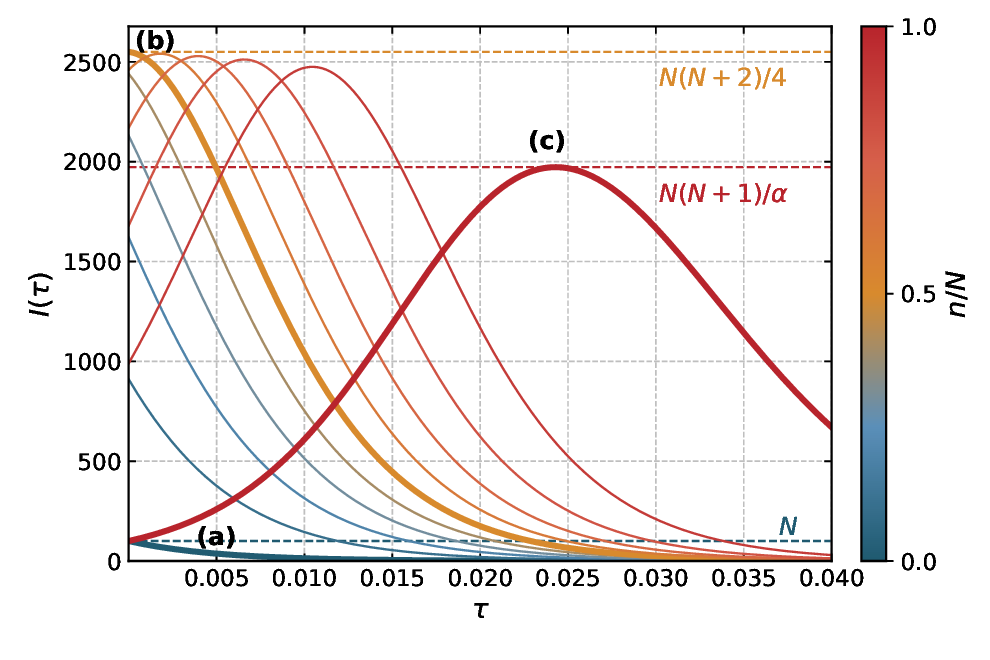}
    \caption{Intensity of the Dicke states $|j, n-j\rangle$ for $N=100$, with bold lines for $n=1 (a),50 (b), 100 (c)$ to highlight the low, central, and high excitation regimes. Dashed lines were added for the peak of the pulse to guide the eye.
    The peak intensity of \eqref{Dicke_excited} is given by the asymptotic numerical formula $I=\frac{N(N+1)}{\alpha}$, where $\alpha = 5.12$. The colorbar shows the excitation fraction.
    }
    \label{fig:W}
\end{figure}

\noindent
Dicke states are symmetric collective-spin eigenstates in the $j=N/2$ manifold. Three salient states are the all-excited, $|e\rangle_\ell = \begin{pmatrix}1\\0\end{pmatrix}$, all-ground, $|g\rangle_\ell = \begin{pmatrix}0\\1\end{pmatrix}$, and central even-N eigenstates.
Of these three, only the third cannot be written as a product state, and is therefore entangled:
\begin{subequations}
\begin{align}
|j,j\rangle 
&= |e,\dots,e\rangle
= \bigotimes_{\ell=1}^{2j} |e\rangle_\ell ,
\label{Dicke_excited}
\\
|j,-j\rangle
&= |g,\dots,g\rangle
= \bigotimes_{\ell=1}^{2j} |g\rangle_\ell ,
\label{Dicke_ground}
\\
|j,0\rangle
&= \binom{2j}{j}^{-1/2}
\sum_{\mathfrak{P}} 
|e_1,\dots,e_j,g_{j+1},\dots,g_{2j}\rangle.
\label{Dicke_central}
\end{align}
\end{subequations}
where the sum $\sum_\mathfrak{P}$ adds all distinct permutations. 
Among the three, only $|j,0\rangle$ exhibits superradiance from the start.
The ground-state \eqref{Dicke_ground} cannot undergo spontaneous emission, as it is a dark state.
Low-excitation Dicke states, exemplified by the singly-excited state (often called the $W$-state \cite{DurVidalCirac2000Wstates}) in figure \ref{fig:W}(a) peak initially and decay immediately.
By contrast, states initialized sufficiently high on the Dicke ladder build up their intensity as their populations descend toward transitions with larger decay rates, as illustrated by the all-excited state in figure \ref{fig:W}(c).
Since the initial intensity is equal to $\frac{\lambda_n}{2\Gamma}=n(N-n+1)$, the difference between the intensity of the transition from level $n$ to $n-1$:
\begin{equation}
\frac{\lambda_{n-1}-\lambda_n}{2\Gamma}=2n-N-2,
\end{equation}
indicates that the intensity initially builds up only when $n>(N+2)/2$.

The CDS \eqref{Dicke_central} and $|j,1\rangle$ states are highly entangled and have the largest initial intensity for even $N$. 
Indeed, we notice that for the total intensity:
\begin{equation}
I(0)=\langle j,m|\hat J_+\hat J_-|j,m\rangle =\Big[j(j+1)+\frac{1}{4}-\big(m-\frac{1}{2}\big)^2 \Big].
\end{equation} 
This tells us that the greatest initial intensity happens for $m=1/2$ for an odd number of atoms $\big(I(0)=J^2+1/4\big)$; 
and $m=0,1$ for even $N$ $\big(I(0)=J^2\big)$.

Let us express the Dicke states in the excitation number representation, so that $|j,m\rangle \equiv |n-j\rangle$.
For the CDS ($|0\rangle$), the initial population is $P_n(0)=\delta_{n,j}$, and therefore only the level $n=j$ is initially occupied.
Since collective decay can only lower the excitation number, the subsequent dynamics is confined to $0\leq n \leq j$.
The populations are then given by \cite{Agarwal_Master_Equation,alabbar2026superradiant}:
\begin{align}
P_n(\tau) = C_{n,j}(\tau)
=
\left[\displaystyle\prod_{r=n+1}^{j}\frac{\lambda_r}{2\Gamma}\right]
\sum_{m=n}^j
\frac{e^{-\lambda_m\tau/\Gamma}}{\displaystyle\prod_{\substack{\ell=n\\ \ell\neq m}}^{j}\frac{\lambda_\ell - \lambda_m}{2\Gamma}}.
\label{Dicke_Populations}
\end{align}
The factors $\frac{1}{2\Gamma}$ in the products cancel exactly, and may be dropped. 
Although the decay rates satisfy the degeneracy $\lambda_n=\lambda_{N-n+1}$, the levels above $n=j$ are never occupied during the decay of the CDS.
Consequently, each decay path contains only one member of every degenerate pair, so all poles remain simple and no polynomial-exponential terms appear. 
Such terms arise in the general transfer coefficients $C_{n,k}(\tau)$ when the initial excitation number satisfies $k>j$, because the decay path can then encounter both degenerate decay rates.

We finally notice that $P_n(\tau)$ is simply the transfer coefficients $C_{n,j}(\tau)$ which is the probability that a system starting in state $j$ has decayed into state $n$ by time $\tau$.
Since the dipole moment of eq. \eqref{Dipole_Moment_analytical} vanishes for all $n$, the coherent intensity contribution $D^2(\tau)=|\langle \hat J_+\rangle|^2(\tau)=0$, and thus the total intensity $I$ is incoherent $\mathcal{I}$: 
\begin{equation}
\mathcal{I}(\tau)=
I(\tau) -D^2(\tau)=I(\tau)=
\sum_{n=0}^{j}\frac{\lambda_n}{2\Gamma} C_{n,j}(\tau)
.
\end{equation}

\begin{comment}
\begin{figure}
    \centering
    \includegraphics[width=1\linewidth]{figures/Figures_All/fig3_dicke_g2.eps}
    \caption{The Glauber function for Dicke states, calculated for $N=100$ with $n=2,50, 100$ excitations. The lines have been cut after the intensity reaches $10\%$ of its peak value: For continuity, that is where the dotted lines begin. Reference black dashed lines at $g^{(2)}(\tau)=1$ were added for the Poissonian limit.}
    \label{fig:Dicke_g2}
\end{figure}
\end{comment}

\subsection{Mean-Field Solution}
\noindent Consider a Dicke state with an excitation fraction $x_0\equiv n/N\approx 1/2$. 
We use the density functions $W(x,0)=\delta(x-x_0)$ and $W(x,\tau)=\delta(x-x(\tau))$. 
Then, the drift velocity will be $d(x)=\frac{dx}{d\tau}=-2Nx(a-x)$. 
We separate the variables and integrate both sides to get:
\begin{equation}
x(\tau)=\frac{ax_0e^{-2(N+1)\tau}}{a-x_0(1-e^{-2(N+1)\tau})},
\end{equation}
where $a=1+1/N$, just as in \eqref{Lambda_x}. 
Finally, with:
\begin{equation}
\frac{A \cosh{z}+B\sinh{z}}{\sqrt{A^2-B^2}}=\cosh{\left(z+\text{arctanh}\big\{B/A\big\}\right)},
\label{Hyperbolic_Function_Property}
\end{equation}
the mean-field intensity is found from:
\begin{align}
I_{\text{MF}}(\tau)
&\approx
\frac{1}{2\Gamma} \int_0^1 dx \lambda(x) W(x,\tau)
=\frac{\lambda(x(\tau))}{2\Gamma}
\notag \\
&=
\frac{(N+1)^2 x_0 (a-x_0)}{\left( ae^{(N+1)\tau}-2x_0 \sinh((N+1)\tau) \right)^2}
\\
&=
\frac{(N+1)^2}{4}\text{sech}^2\{(N+1)(\tau-\tau_D)\}
,
\label{Dicke_MF_1}
\end{align}
where $\tau_D = \frac{1}{2(N+1)}\ln\left( \frac{2\Gamma n^2}{\lambda_{n}} \right)$ is the formal pulse-delay.

\begin{figure}
    \centering
    \includegraphics[width=1\linewidth]{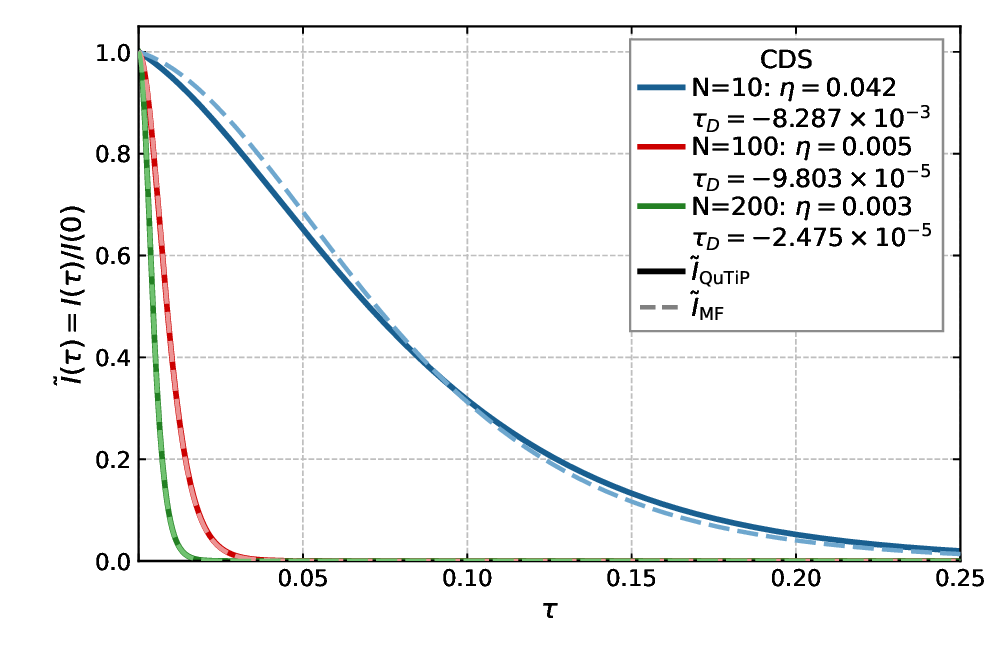}
    \caption{The CDS Intensity comparison.}
    \label{fig:Dicke_Mean_FP}
\end{figure}

Next, from this solution, we extract the pulse-width:
\begin{equation*}
Na\Delta\tau_{1/2}=\text{arcsech}\Big\{1/\sqrt{2}\Big\}
=\ln(1+\sqrt{2}),
\end{equation*}
where $\Delta \tau_{1/2}=\tau_{1/2}-\tau_D$ is the descending half-width at half maximum.
This approximation is most accurate near the center of the Dicke ladder, and becomes less (but still reasonably) accurate toward the low- and high-excitation edges where the pulse need not retain a symmetric $\text{sech}^2$ profile.
In particular, low-excitations exhibit approximately exponential decay as shown in \ref{fig:W} (a).
Nevertheless, for the descending tail, we may approximate $\text{sech}^2z\approx 4e^{-2z}$ to obtain $I(\tau)\approx (N+1)^2e^{-2(N+1)(\tau-\tau_D)}$. 
Furthermore, for the CDS, $\tau_D = \frac{1}{2(N+1)}\ln(N/(N+2))<0$. 
This negative value should \textit{not} be interpreted as a pulse-peak occurring before the state was even prepared!
The actual reason for the backward shift is due to the finite-$N$ continuum rate whose maximum occurs at $x_{\text{peak}}=\frac{a}{2}=\frac{1}{2}+\frac{1}{2N}$.
The resulting offset produces $\tau_D\sim-1/N^2$, which vanishes as $N\rightarrow \infty$ (where diffusion can be appropriately ignored).
For $N=100$, $\tau_D\approx -9.80\times 10^{-5}$.
Fig. \ref{fig:Dicke_Mean_FP} presents the results.

\section{Eigenstates of $\hat J_x$: \\ Rotated Dicke States}
\subsection{Analytical Solution}
\noindent 
Dicke states have no macroscopic transverse dipole, so their cooperative spontaneous emission contains no coherent contribution and is initiated by quantum fluctuations. 
Glauber and Haake showed this explicitly for the fully inverted state; although its mean polarization and the incident electric field both vanish, the polarization cannot be sharp because it does not commute with the inversion, and its nonzero initial fluctuations seed the first photons that are subsequently amplified by the cooperative dynamics \cite{GLAUBER197829}.
The quantum vacuum is the ground state of the electromagnetic modes, with zero mean field but nonzero field correlations.
Its coupling to the atoms supplies the quantum noise accompanying radiative damping, and therefore, vacuum fluctuations and radiation reaction are complementary descriptions of the same fluctuation--dissipation process \cite{milonni2019introduction}. 

A spatially uniform resonant microwave, radio-frequency, or Raman pulse can rotate all atomic pseudo-spins together relative to the fixed decay axis, thereby converting the longitudinal population imbalance of a Dicke state into a transverse collective dipole \cite{Lucke2014,Toyoda2011}. 
Because a collective rotation leaves the state’s entanglement unchanged and alters only its orientation relative to the decay axis, these rotated states reveal:
(i) how the orientation alone changes the pulse profile and photon statistics; 
(ii) how the induced transverse dipole redistributes the total intensity between coherent and incoherent emission;
(iii) how that dipole now determines the initial decay rate; and,
(iv) how the states of this section connect to those yielded from transferring the squeezing from the bath of correlated bosonic fields to the atoms, and to the ACS.

For the purposes of this article, we will call the eigenstates of the $\hat J_x$ operator with 
$\hat J_x|n-j\rangle_x=(n-j)|n-j\rangle_x$, 
the Rotated Dicke States (RDS).
Since 
$\hat J_x = e^{-i \frac \pi 2 \hat J_y} \hat J_z e^{+i\frac\pi2\hat J_y}$, 
where 
$\hat R_y(\pi/2)\equiv e^{-i \frac \pi 2 \hat J_y}$,
we can write 
$|n-j\rangle_x \equiv \hat R_y(\pi/2)|n-j\rangle$.

We expand the RDS in terms of Dicke states using the Wigner small-d matrix elements \cite[eq. (3.203)]{Sakurai_Napolitano_2020}:
\begin{equation}
|0\rangle_x =\sum_{n=0}^{N} 
d_{n-j,0}^{(j)}\left(\frac\pi2\right)
|n-j\rangle,
\end{equation}
Note that $d_{n-j,0}^{(j)}\left(\frac\pi2\right)=0$ for odd $n$.
As for even $n$:
\begin{align}
d^{(j)}_{n-j,0}\left(\frac{\pi}{2}\right)
&=
\frac{j!}{2^j}
\sum_{\ell}
(-1)^{n-j+\ell}
\notag \\
&\times \frac{\sqrt{(n)!(N-n)!}}
{(j-\ell)!(N-n-\ell)!\,\ell!\,(n-j+\ell)!}
\notag \\
&= \frac{(-1)^{n/2}}{2^j}
 \sqrt{\binom{n}{n/2}\binom{N-n}{(N-n)/2}}.
\end{align}
where $\ell\in \mathbb Z$ such that 
$\max\{0,j-n\}\leq\ell\leq\min\{j,N-n\}$. 
Then, the initial population of the even $n^{\text{th}}$ level is:
\begin{equation}
P_n(0) =
\left| d^{(j)}_{n-j,0}\left(\frac{\pi}{2}\right)
\right|^2
=
\frac{n! (N-n)!}{2^N 
\left[\Big(\frac{n}{2}\Big)!\Big(\frac{N-n}{2}\Big)!\right]^2}.
\label{RDS_pop}
\end{equation}
However, since the central RDS now also occupies levels $k>j$, its decay paths can contain both members of the degenerate decay rates.
The transfer coefficients therefore contain repeated poles. 
This is why the simple-pole expression used for the CDS is insufficient, and we use the repeated-pole formula in Eq. \eqref{Repeated_pole_transfer}.

The intensity for the central RDS can be found from the transfer coefficients $C_{n,k}(\tau)$ as defined in the appendix. 
For a decay path from $k$ to $n$, with $0\leq n \leq k \leq N$, we rewrite the rate multiplicity from eq. \eqref{multiplicity_A}:
\begin{equation}
m_{n,k}(p)
=
\sum_{q=n}^{k}\delta_{\lambda_q,\lambda_p},
\qquad n\le p\le k,
\label{multiplicity_B}
\end{equation}
where the Kronecker delta $\delta_{\lambda_q,\lambda_p}=1$ if $q=p$ or $q=N-p+1$, and vanishes otherwise. 
We also define the reduced pole denominator and the reduced reciprocal pole sum, respectively given by:
\begin{equation*}
S^{(1)}_{n,k}(p) =
\prod_{\substack{q=n\\ \lambda_q\neq \lambda_p}}^{k}
(\lambda_q-\lambda_p);
\quad 
S^{(2)}_{n,k}(p) = \sum_{\substack{q=n\\ \lambda_q\neq \lambda_p}}^{k}
\frac{1}{\lambda_q-\lambda_p},
\end{equation*}
in the convention that empty products are taken to be one, and empty sums are taken to be zero.

\begin{figure}
    \centering
    \includegraphics[width=1\linewidth]{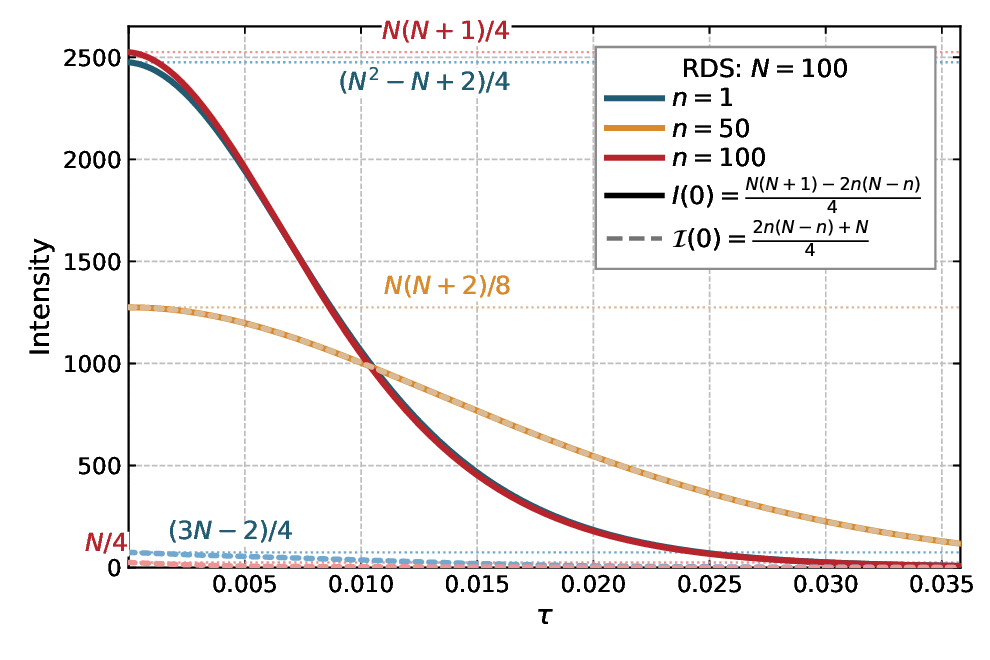}
    \caption{The total and incoherent intensity of RDS.}
    \label{fig:RDS_I}
\end{figure}

Since $\lambda_p=\lambda_{N-p+1}$, the transfer coefficients may contain second-order poles. 
For the states considered here, no pole is higher than second order, and therefore the inverse-Laplace transform can be written as:
\begin{align}
C_{n,k}(\tau)
&=
\left(
\prod_{r=n+1}^{k}\lambda_r
\right)
\Bigg[
\sum_{p=n}^{k}
\frac{\bigl(2-m_{n,k}(p)\bigr)}{S^{(1)}_{n,k}(p)}e^{-\frac{\lambda_p\tau}{\Gamma}}
\label{Repeated_Pole_Transfer_Explicit}
\\
&+
\frac{1}{2}
\sum_{p=n}^{k}
\frac{\bigl(m_{n,k}(p)-1\bigr)}{S^{(1)}_{n,k}(p)}
\left(
\frac{\tau}{\Gamma}
-
S^{(2)}_{n,k}(p)
\right)
e^{-\frac{\lambda_p\tau}{\Gamma}}
\Bigg].
\notag
\end{align}
And then, the initial intensities follow: 
\begin{subequations}
\begin{align}
&4I(0)=N(N+1)-2n(N-n);
\label{RDS_Intensity}
\\
&4\mathcal I(0)=2n(N-n)+N=4\Big\langle \big(\Delta \hat J_z\big)^2\Big\rangle.
\label{RDS_Incoherent}
\end{align}
\end{subequations}
Their difference gives the coherent contribution $D^2(0)=(n-j)^2=m^2$.
The central RDS maximizes $\mathcal I(0)$ and minimizes $I(0)=\frac{N(N+2)}{8}=\frac12I^{\text{CDS}}(0)$.
Moving toward either the low- or high-excitation edges reduces the fluctuation contribution of the pulse.
States with $n$ or $N-n$ initial excitations have identical total and incoherent intensities, but opposite dipole amplitudes, so intensity measurements alone cannot distinguish their relative phase.
Accordingly, the vertical separation between the solid and dotted curves in Fig. \ref{fig:RDS_I} directly displays the portion of the radiation carried by the collective dipole.

The origin of the initial rise or fall of the intensity follows directly from the population dynamics:
\begin{equation*}
\partial_\tau I(\tau)=4\Big \langle\hat J_+\hat J_z\hat J_- \Big \rangle
=\frac{2}{2\Gamma}\sum_{n=1}^N 
P_n(\tau)\lambda_n \left(\frac{\lambda_{n-1}-\lambda_n}{2\Gamma}\right),
\end{equation*}
meaning that each occupied level contributes its present emission rate $\lambda_n/2\Gamma$ weighted by the change in the rate available after its decay to $n-1$.
But since $\left(\frac{\lambda_{n-1}-\lambda_n}{2\Gamma}\right)=2(n-j-1)$, we have three cases:
(i) Levels with $n>j+1$ feed transitions with larger rates and contribute to a buildup;
(ii) levels with $n<j+1$ feed transitions with smaller rates and contribute to a decay; and,
(iii) levels with $n=j+1$ contribute nothing to the slope.
For the $\hat J_z$-basis population distribution fixed by the $\hat J_x$-eigenvalue condition, this rate-weighted population imbalance is always nonpositive, yielding the identity:
\begin{equation}
\partial_\tau I(\tau)|_{\tau=0}
=-4D^2(0)
=-4(n-j)^2\leq 0,
\label{RDS_Caveat}
\end{equation}
which gives a negative slope for every noncentral RDS, and vanishes for the central one.
For the latter case, we see that there is a stationary maximum at $\tau=0$:
\begin{equation}
\partial_\tau^2I(\tau)|_{\tau=0}
=-J^2(J^2+6).
\label{RDS_caveat2}
\end{equation}
Thus, both the central and noncentral RDS peak immediately but for different reasons.
For the former, the upward and downward changes of the symmetric population distribution cancel at first order, after which depletion of the rapidly emitting central levels drives the intensity downward.
For the latter, it is because their pre-existing dipoles begin to decay at once.

Since Dicke states have highly localized excitations, their phase-uncertainty is undefined.
However, the initial phase uncertainty for the central RDS scales as 
$\Delta \phi = \sqrt{\frac{2}{N(N+2)}} \sim 1/N$ 
within the estimate used here, demonstrating that the RDS achieve Heisenberg-limited phase sensitivity at $\tau = 0$.
This highlights the metrological utility of the RDS as a highly squeezed state along the phase coordinate.
Nevertheless, as collective superradiant decay progresses, collective uncoupling and quantum cascade fluctuations rapidly destroy this coherence, causing the phase uncertainty to grow monotonically.

\begin{figure*}
    \centering
    \includegraphics[width=1\linewidth]{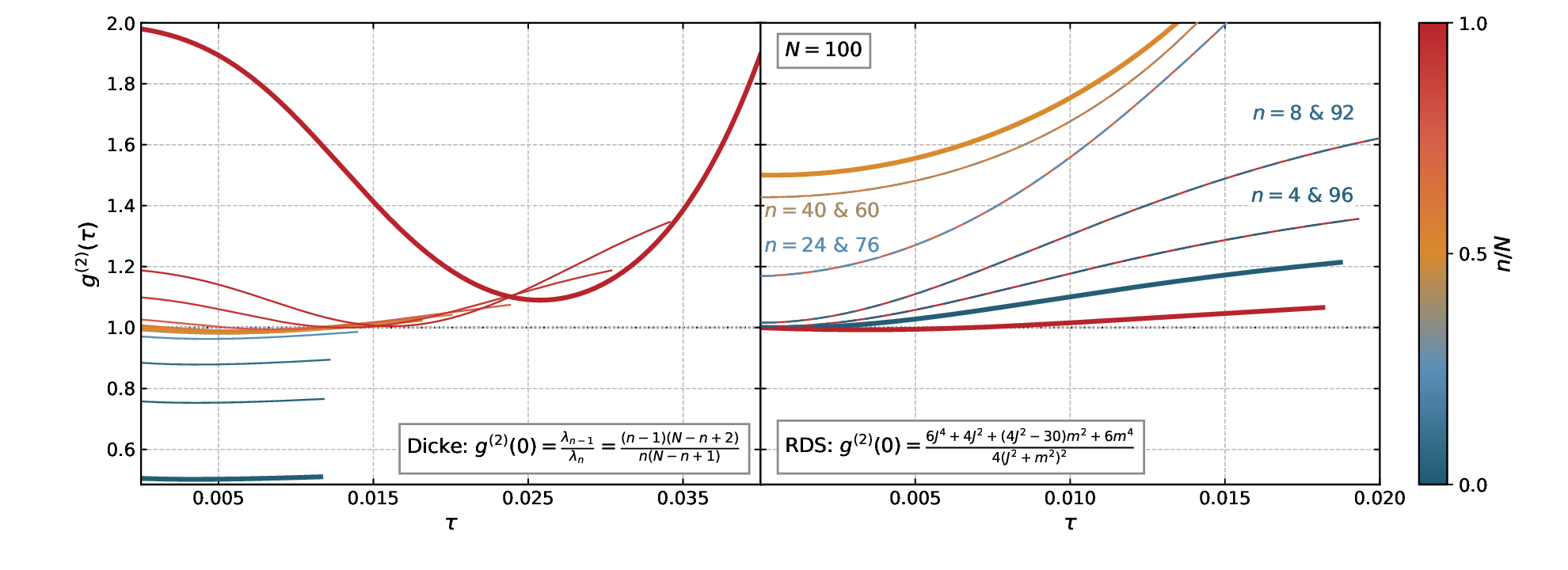}
    \caption{The Glauber function for some representative levels. For RDS, dashed lines show degenerate curves.}
    \label{fig:RDS_g2}
\end{figure*}

Figure \ref{fig:RDS_g2} displays the Glauber function for Dicke states and RDS, with $n\in\{2,50,100\}$ in bold, and the pairs 
$n\in\{(4,96), (8,92), (24,76), (40,60)\}$
in thinner lines.
The initial scale is displayed for both states.
In the figure, we see that the Dicke states display three distinct behaviors:
The $n=2$ state begins antibunched at $g^{(2)}(0)\approx 0.5$ and rises slowly;
the CDS begins nearly Poissonian and increases only modestly;
and the fully excited state begins near $g^{(2)}(0)=2$, falls toward the Poissonian value, and subsequently rises.
The RDS instead satisfies the symmetries 
$I_m(\tau)=I_{-m}(\tau)$ and $g_m^{(2)}(\tau)=g_{-m}^{(2)}(\tau)$,
because of the Liouvillian covariance of the $\hat U=e^{-i\pi\hat J_z}$ rotation operator 
$\mathcal L\{\hat U\rho\hat U^\dagger\}=\hat U\mathcal{L}\{\rho\}\hat U^\dagger$,
a discrete $\mathbb{Z}_2$ symmetry under a rotation by $\pi$ about the $z$-axis \cite{Ramond2010}.
For the central RDS, 
\begin{subequations}
\begin{align}
&g^{(2)}(0)=\frac{3}{2}+\frac{1}{J^2};
\\
\partial_\tau &g^{(2)}(0)=-2\left(1+\frac{6}{J^2}\right)<0.
\end{align}
\end{subequations}
Thus, although it is initially bunched, its normalized pair correlation immediately decreases.
Since $\partial_\tau I(0)=0$, this initial decrease arises from the reduction of $\langle \hat J_+^2\hat J_-^2 \rangle$.
The curve reaches a shallow minimum at a small positive time and then rises over the remainder of the emission interval.
Moreover, the Glauber function is monotonically decreasing with $m^2\leq j^2<J^2$:
\begin{equation}
\partial_{m^2}g^{(2)}(0)=\frac{(4J^2+15)m^2-J^2(4J^2+19)}{2(J^2+m^2)^3}<0.
\end{equation}
We see from this that the central RDS combines a vanishing dipole with the strongest initial normalized pair correlations within the RDS ladder.

\subsection{Mean-Field Solution}
\begin{figure}
    \centering
    \includegraphics[width=0.96\linewidth]{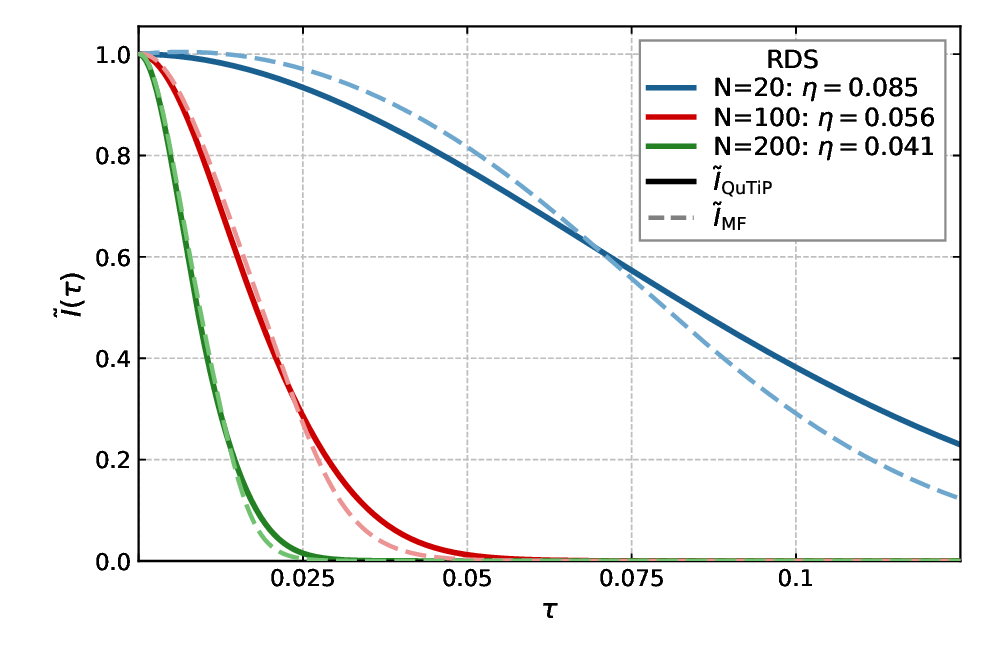}
    \caption{The RDS Intensity comparison.}
    \label{fig:ItRDS}
\end{figure}

\noindent The decay shape of the continuum can be determined approximately from the mean-field approach. 
We see that $\lambda(x)=2\Gamma N^2 x(a-x)$, and from \eqref{RDS_pop} and Stirling's approximation $\Big(n! \approx \sqrt{2\pi n}\left(\frac{n}{e}\right)^n\Big)$, we can simplify the population for large-$N$:
\begin{equation}
P_n(0)=
\frac{4^{\frac{n}{2}}4^{\frac{N-n}{2}}}{2^N\sqrt{\frac{\pi}{2}n}\sqrt{\frac{\pi}{2}(N-n)}}
=
\frac{2}{\pi\sqrt{n(N-n)}},
\end{equation}
and $W(x,0)=jP_n(0)=\frac{1}{\pi\sqrt{x(1-x)}}$. 
Then, we find that $x(\tau;x_0)=\frac{ax_0 e^{-2(N+1)\tau}}{a-x_0(1-e^{-2(N+1)\tau})}$ by solving the drift equation. 
If we define the mean-field propagation parameter:
\begin{equation}
\varkappa(N,\tau)=\frac{\sqrt{a}-\sqrt{a-1+e^{-2(N+1)\tau}}}{\sqrt{a}+\sqrt{a-1+e^{-2(N+1)\tau}}}
,
\label{mean_field_propagator}
\end{equation} 
where for large-$N$, we get $\varkappa\xrightarrow{N\gg1}\tanh\left(j\tau\right)$. 
Then:
\begin{align}
I_{\text{MF}}(\tau)
&\approx
\frac{1}{2\Gamma} \int_0^1 dx_0 \lambda(x(\tau;x_0))W(x_0,0) 
\label{Mean_Field_Intensity_2}
\\
&=\frac{(Na)^2}{\pi}\int_0^1\frac{dx_0}{\sqrt{x_0(1-x_0)}}\,\frac{ e^{-2(Na)\tau}x_0(a-x_0)}{\left(a-x_0(1-e^{-2Na\tau})\right)^2}
\notag \\
&=
\frac{N^2}{8}
\frac{1+\varkappa}{(1-\varkappa)^3}
\left[
(1-\varkappa)^2-\frac{4\varkappa}{N}
\right]
\left[
(1-\varkappa)^2+\frac{4}{N}
\right]
\notag\\
&\xrightarrow{N\gg1}
\frac{N^2}{8}\text{Sech}^2(j\tau)
,
\end{align}
where the Stieltjes transform \cite{doi:10.1073/pnas.23.4.242} was used:
\begin{subequations}
\begin{align}
&\mu(dx)=\frac{dx}{\pi \sqrt{x(1-x)}}
\label{Stieltjes_Measure}
;\\
&
S_0(\alpha, \beta) = \int_0^1 \frac{\mu(dx)}{\alpha-\beta x}
=
\frac{1}{\sqrt{\alpha(\alpha-\beta)}}
\label{Stieltjes_Transform_Order_0}
;\\
&S_1(\alpha, \beta) = \int_0^1 \frac{x\mu(dx)}{\alpha-\beta x}
=
\frac{\alpha}{\beta } S_0(\alpha,\beta)
-\frac{1}{\beta}.
\label{Stieltjes_Transform_Order_1}
\end{align}
\end{subequations}
This MF approximation overshoots the exact results by $2/(N+2)$.
Fig. \ref{fig:ItRDS} compares the normalized emission intensity from the full quantum dynamics with the mean-field prediction in Eq. \eqref{Mean_Field_Intensity_2}.
The fit improves with increasing system size, with the NRMSE decreasing from $0.13$ at $N=20$ to $0.05$ at $N=200$, consistent with the expected large-$N$ validity.
Finally, the MF pulse-width for the central RDS is $\tau_{1/2}^{\text{RDS}}\approx2a\Delta\tau_{1/2}^{\text{CDS}}$.
\section{States from the Squeezed Bath:
\\
Squeezed Dicke States
}
\subsection{The Steady State Solution}
\noindent Next, we consider a class of states which we will call Squeezed Dicke states (SDS). 
These states are of interest to us, because the bath squeezing continuously deforms the central RDS population distribution, providing a reservoir-controlled test of how pair correlations determine the collective-emission pulse shape and photon statistics.
Typically, the squeezed bath is realized by parametric amplifiers over a broad range of frequencies \cite{Wu1986, Qiu2023, Juanes2026}.

Agarwal and Puri \cite{AGARWAL1989267,PhysRevA.41.3782} have shown that the SDS can be produced from the interaction of qubits with a broadband squeezed bath (SB), with a squeezing parameter $r$ such that each correlated mode pair in the bath has a mean photon occupation $\bar n=\sinh^2r$ and an anomalous pair correlation $M=e^{i\phi}\sqrt{\bar n(\bar n+1)}=\frac{e^{i\phi}}{2}\sinh(2r)$, where $\phi$ is the squeezing phase.
Let us choose the transverse phase origin $\hat J_\pm \rightarrow e^{\pm i\phi/2}\hat J_\pm$, so that the anomalous correlation is real and positive, and work in the interaction frame in which the SB correlations are stationary. 
Then, the SB master equation is given by \cite{PhysRevA.41.3782}:
\begin{equation}
\label{Squeezed_Bath_ME}
\begin{aligned}
\mathcal{L}_{\textbf{SB}}^{\{r\}}\{\rho(t)\}
&={}
\Gamma\cosh^2r
\Bigl(
2\hat J_- \rho \hat J_+
-\{\rho, \hat J_+ \hat J_-\}
\Bigr)
\\[0.3em]
&+\frac{\Gamma}{2}
\sinh2r  
\Bigl(
2\hat J_+ \rho \hat J_+
-\{\rho, \hat J_+ \hat J_+\}
\Bigr)
\\[0.3em]
&+\frac{\Gamma}{2}
\sinh2r  
\Bigl(
2\hat J_- \rho \hat J_-
-\{\rho, \hat J_- \hat J_-\}
\Bigr)
\\[0.3em]
&+\Gamma \sinh^2r
\Bigl(
2\hat J_+ \rho \hat J_-
-\{\rho, \hat J_- \hat J_+\}
\Bigr),
\end{aligned}
\end{equation}
whose steady states are obtained by applying a nonunitary excitation-number filter to the central RDS:
\begin{equation}
    |\Psi_0\rangle 
    = \frac{\exp\Big\{\delta\big(j+\hat J_z\big)\Big\}}{\sqrt{\mathfrak{Q}_0(e^{4\delta})}} e^{-i\frac{\pi}{2} \hat J_y} |0\rangle
    \equiv |0\rangle_x^\delta, 
\end{equation}
with $\delta=\ln\sqrt{\tanh r}$, and the moment function, $\mathfrak{Q}$, is defined from:
\begin{equation}
\mathfrak{Q}_\ell (z) 
=\left(z\partial_z\right)^\ell \mathfrak{Q}_0(z)=
\sum_{k=0}^j k^\ell z^k
\Big| d_{2k-j,0}^j \Big(\frac{\pi}{2}\Big)\Big|^2.
\label{Q_function}
\end{equation}
These steady states of \eqref{Squeezed_Bath_ME} are the vacuum states of the operator \cite{AGARWAL1989267,PhysRevA.41.3782,zhang2025time}:
\begin{equation}
\Big(
\hat J_- \cosh r+\hat J_+ \sinh r
\Big) |\Psi_0\rangle=0.
\end{equation}

We may expand the state $|0\rangle_x^\delta$ in terms of Dicke states $|n-j\rangle$ with amplitudes $c_n$ to find that they vanish for odd $n$, but for even $n$ \cite{PhysRevA.41.3782}:
\begin{equation}
c_n =  
\frac{e^{n\delta} d^{(j)}_{n-j,0}\left(\frac{\pi}{2}\right)}{\sqrt{\mathfrak{Q}_0(e^{4\delta})}},
\label{amplitudes_SSB}
\end{equation}
and $P_n(0;\delta)=|c_n|^2$. 

At $\tau=0$, the prepared state is removed from the bath and evolves under the vacuum Liouvillian.
The superradiant pulse is highly dependent on the initial squeezing parameter. 
Using $r$ as a knob, we realize that as it increases, the atomic system comes to resemble the RDS. 
In fact, we can define a crossover between the shape of the superradiant intensity being that of the weakly squeezed bath, to a fully saturated system $r\rightarrow\infty$ that becomes the RDS (when $\tanh(r)\rightarrow 1$). 
The relation between the populations is found to be:
\begin{equation}
P_n(0;\delta)=\frac{e^{2n\delta}}{\mathfrak{Q}_0(e^{4\delta})}
P_n^\text{RDS}(0)
,
\label{Populations_SSB}
\end{equation}
where $P_n^\text{RDS}(0)=\left| d^{(j)}_{n-j,0}\left(\frac{\pi}{2}\right)
\right|^2$. 
To find the intensity plotted as the solid curves of Fig. \ref{fig:SSB_MF_I}, we recognize first a property of the moment function (and accounting for the pair-excitation structure, we have $k=n/2$):
\begin{equation}
\frac{\mathfrak{Q}_\ell(e^{4\delta})}{\mathfrak{Q}_0(e^{4\delta})}
=
\sum_{k=0}^j (k)^\ell P_k(0;\delta)=\langle k^\ell\rangle.
\end{equation}
Then, for the total intensity, for example, we have $\langle \lambda_n/2\Gamma\rangle=\langle n(N-n+1)\rangle =2(N+1)\langle k\rangle-4\langle k^2\rangle$, and so:
\begin{equation}
I(0; \delta) 
= \frac{2(N+1)\mathfrak{Q}_1(e^{4\delta})-4\mathfrak{Q}_2(e^{4\delta})}{\mathfrak{Q}_0(e^{4\delta})}.
\label{SSB_Intensity_Q}
\end{equation}
The formula for the Glauber function for the central SDS is given in Fig. \ref{fig:SSB_g2}.

For the noncentral SDS, the parity structure is different. Let us first write the population imbalance $Z_m(0)=\,_x^\delta\langle m|\hat J_z|m\rangle_x^\delta$ to find our dynamical equations \eqref{Dipole_Moment_analytical}--\eqref{Glauber_function_Analytical}:
\begin{subequations}
\begin{align}
D(0)
&=|m|\sech\delta;
\label{Dipole_SSB}
\\
\mathcal I(0)
&=-2\sinh^2r Z_m(0).
\label{Noncentral_SSB_Intensity}
\end{align}
\end{subequations}
and the Glauber function is given in a similar method to $\ref{fig:SSB_g2}$, but with 
$\mathfrak{Q}^{m}_\ell(e^{2\delta})=\sum_{n=0}^N n^\ell e^{2n\delta} 
\left| d^{(j)}_{n-j,m}\left(\frac{\pi}{2}\right)\right|^2$.

\subsection{Squeezing-Controlled Crossover}
\noindent
Now, it can be seen that the SDS are obtained by exponentially reweighting the central RDS amplitudes according to their excitation number. 
Then, to any precision, we can see how saturated the states are, and how close we are to the RDS through the crossover $r_c = \text{arctanh}\{e^{2\delta_c}\}$. 

First, let us define the intensity ratio:
\begin{equation}
R(N,r)=\frac{I_\text{SDS}\big(0;\ln\sqrt{\tanh r}\big)}{I_\text{RDS}(0)}.
\label{Intensity_Ratio}
\end{equation}
Furthermore, the moment function is related to the Legendre polynomials through:
\begin{equation}
\mathfrak{Q}_0(z)=z^{j/2}\mathcal{P}_j\left(\frac{1+z}{2\sqrt{z}}\right).
\end{equation}
If we take the large-$N$ limit, the Mehler--Heine approximation \cite[Th. 8.1.1]{szego1939} gives us:
\begin{equation}
\lim_{N\rightarrow \infty} \mathcal P_j(\cosh(2\delta))=\mathfrak{I}_0(N\delta),
\notag
\end{equation} 
where \cite{osti_243992}:
\begin{equation}
\mathfrak{I}_k(z)=\frac{1}{\pi} \int_0^\pi d\theta e^{z \cos \theta} \cos(k\theta)
\end{equation} 
is the modified Bessel function of the first kind of integer-order $k$. 
Then, we see directly that 
$\mathfrak{Q}_0(e^{4\delta})\rightarrow e^{N\delta}\mathfrak{I}_0(N\delta)$. 
Finally, from the definition \eqref{Q_function}, we get:
\begin{align*}
\mathfrak{Q}_1(e^{4\delta})
&\rightarrow \frac{N}{4}e^{N\delta}
\left[
\mathfrak{I}_0(N\delta)+\mathfrak{I}_1
(N\delta)\right];
\\
\mathfrak{Q}_2(e^{4\delta})
&\rightarrow \frac{N^2}{32}e^{N\delta}
\left[
3\mathfrak{I}_0(N\delta)
+
4\mathfrak{I}_1(N\delta)
+
\mathfrak{I}_2(N\delta)
\right].
\end{align*}
Substitute these values into \eqref{SSB_Intensity_Q} with $I_{\text{RDS}}(0)=J^2/2$ into \eqref{Intensity_Ratio} to get the asymptotic limit: 
\begin{equation}
R_\infty(N\delta)= 1-\frac{\mathfrak{I}_2(N\delta)}{\mathfrak{I}_0(N\delta)}.
\label{SSB_Ratio_Asymptotic}
\end{equation} 

\begin{figure}
    \centering
    \includegraphics[width=1\linewidth]{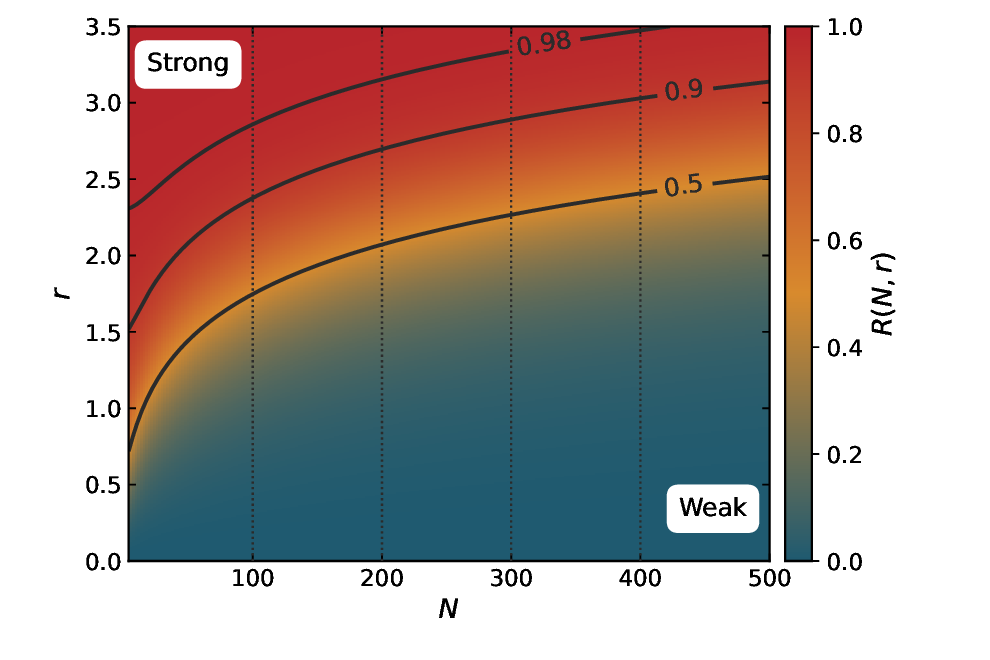}
    \caption{The squeezing crossover diagram for the SDS.
    Three contour lines were added for reference for the intensity ratio $R=50\%, 90\%, 98\%$ from eq. \eqref{Intensity_Ratio}.}
    \label{fig:r_cN}
\end{figure}

\begin{figure}
    \centering
    \includegraphics[width=1\linewidth]{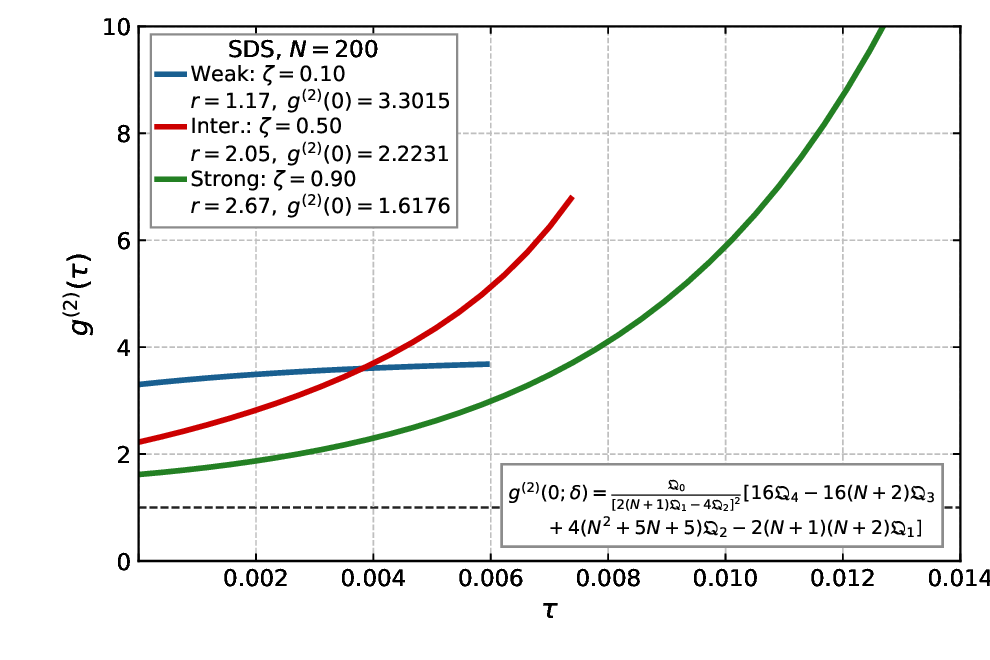}
    \caption{$g^{(2)}(\tau)$ of the central SDS for the three regimes.}
    \label{fig:SSB_g2}
\end{figure}

Now, we define a proximity parameter $\zeta\in[0,1)$ such that $R_\infty(N\delta_c)=\zeta$. 
We can precisely say how close the SDS have come to the RDS limit. 
Since $\delta=\ln\sqrt{\tanh(r)}$, we have $\delta<0$ for finite $r$, $\delta\rightarrow 0^-$ as $r\rightarrow \infty$, such that $N\delta_c$ remains finite. 
Now, let us define the ratio to be found numerically $\delta_c \sim -\frac{a_\infty}{4N}$.
In other words, $a_\infty >0$ is the solution of 
$1-\frac{\mathfrak{I}_2(a_\infty/4)}{\mathfrak{I}_0(a_\infty/4)}=\zeta$.
Then $a_\infty = 4R_\infty^{-1}(\zeta)$, and from the definition of $\delta$, and using 
$e^{-a_\infty/N}\approx1-\frac{a_\infty}{N}$, we get a formula for the crossover:
\begin{equation}
r_c(N;\zeta)= \text{arctanh}\left(e^{-\frac{a_\infty}{2N}}\right)
\approx
\text{arctanh}\sqrt{1-\frac{a_\infty}{N}}.
\label{Crossover}
\end{equation}
For example, if we were to choose $\zeta=90\%$ then $a_\infty =3.842$, and choosing $\zeta=98\%$ yields $a_\infty = 1.622$. 
The phase diagram is displayed in Figure \ref{fig:r_cN}, and the Glauber function is shown for each phase in Figure \ref{fig:SSB_g2}.

\subsection{Mean-Field Approximation}
\noindent Starting from \eqref{Populations_SSB} we have as in the case of RDS only even levels are occupied: $N\Delta x=2$. The initial continuum density will be $W(x,0)=jP_n(0;\delta)$:
\begin{equation}
W(x,0)=\frac{Ne^{2N\delta x}
\left| d^{(j)}_{Nx-j,0}\left(\frac{\pi}{2}\right)
\right|^2
}{2\mathfrak{Q}_0(e^{4\delta})}.
\end{equation}
Let us also take the central-binomial Stirling approximation we used for RDS with 
$\left| d^{(j)}_{Nx-j,0}\left(\frac{\pi}{2}\right)
\right|^2\approx\frac{2}{\pi N\sqrt{x(1-x)}}$ 
and from the large-$N$ limit we used before, we write $\mathfrak{Q}_0(e^{4\delta})\approx e^{N\delta}\mathfrak{I}_0(N\delta)$. 
Then:
\begin{equation}
W(x,0)=\frac{e^{N\delta (2x-1)}}{\pi \mathfrak{I}_0(N\delta)\sqrt{x(1-x)}}.
\end{equation}

From the probability conserving equation, we have $W(x,\tau)=W(x_0,0)|\frac{dx_0}{dx}|$. 
We plug in
\begin{subequations}
\begin{align}
x_0(x,\tau)
&=
\frac{ax}{ae^{-2(N+1)\tau} + x[1-e^{-2(N+1)\tau}]};
\\
\left|\frac{dx_0}{dx}\right|
&=
\frac{x_0^2}{x^2}e^{-2(N+1)\tau},
\end{align}
\end{subequations}
and integrate to get the mean-field intensity:
\begin{align}
I_{\text{MF}}(\tau; \delta)
&\approx
N^2\int_0^1 dx x(a-x)W(x,\tau)
\\
&=N^2\int_0^1 dx_0 x(\tau)[a-x(\tau)]W(x_0,0)
\notag \\
&=
\frac{(N+1)^2e^{-2(N+1)\tau}}{\pi\mathfrak{I}_0(N\delta)}
\notag \\
&\times 
\int_0^1 \frac{dx_0}{\sqrt{x_0(1-x_0)}} 
\frac{e^{N\delta(2x_0-1)}x_0(a-x_0)}
{[a-x_0(1-e^{-2(N+1)\tau})]^2}.
\notag
\end{align}

\begin{figure*}
    \centering
    \includegraphics[width=1.0\linewidth]{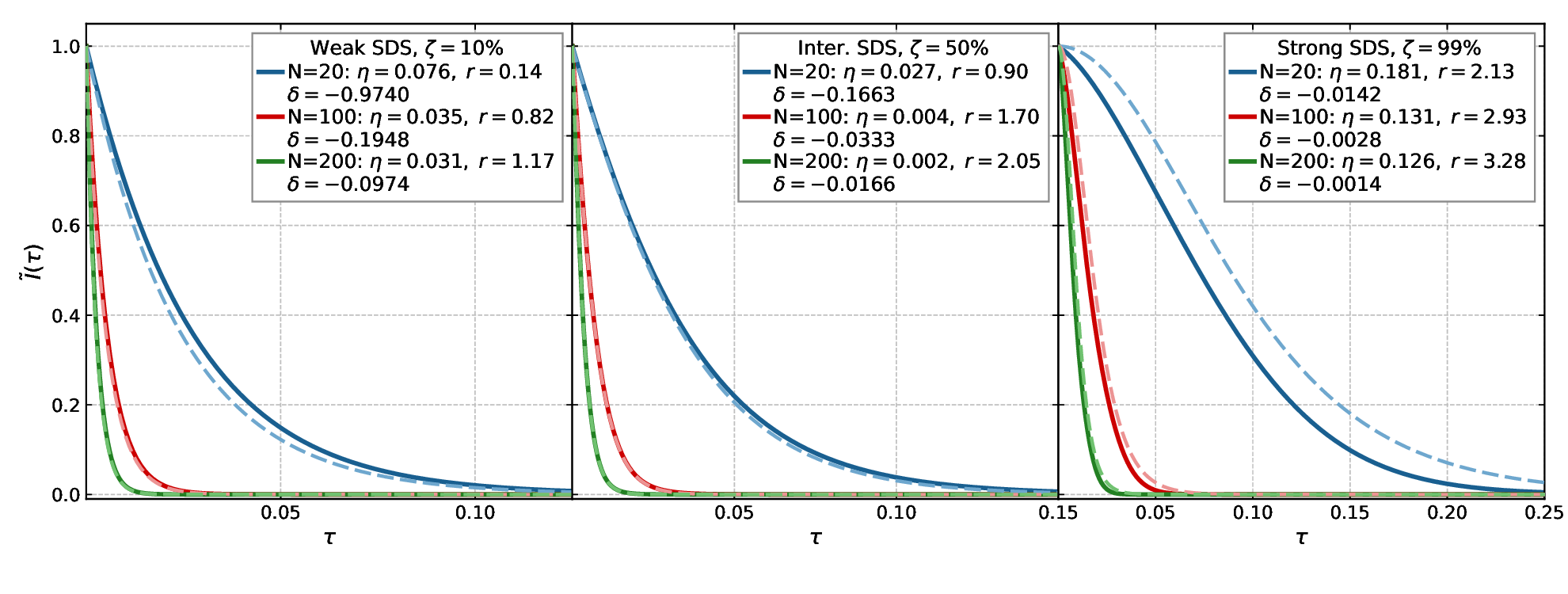}
    \caption{The SDS intensity comparison. 
    The analytical solution (solid) and the $I_{\text{MF}}$ (dashed) were normalized.}
    \label{fig:SSB_MF_I}
\end{figure*}

Let us separate the solutions into the three regimes:
\begin{subequations}
\begin{align}
I_{\text{Strong}}(\tau)
&\xrightarrow[N\gg1]{N|\delta|\ll 1}
\frac{N^2}{8}\text{sech}^2(j\tau);
\label{SSB_I_Strong}
\\
I_{\text{Inter}}(\tau)
&\approx
\frac{N^2}{16}
\frac{(1+\varkappa)^2}{\varkappa^2}
\frac{N(1-\varkappa)^2-4\varkappa}{N\mathfrak{I}_0}
\notag \\
&\times
\Bigg[
\frac{N(1-\varkappa)^2(1+\varkappa^2)+8\varkappa^2}{N(1-\varkappa)^3(1+\varkappa)}
(\mathfrak{I}_0+2\mathfrak{P}_0)
\notag\\
&
-
2\frac{N(1-\varkappa)^2-4\varkappa}{N(1-\varkappa)^2}\mathfrak{P}_1
-\mathfrak{I}_0\Bigg];
\label{SSB_I_Intermediate}
\\
I_{\text{Weak}}(\tau)
&\xrightarrow[N\gg1]{-N\delta\gg 1}
-\frac{N+1}{4\delta}e^{-2(N+1)\tau},
\label{SSB_I_Weak}
\end{align}
\end{subequations}
where $\mathfrak{P}_\ell(\varkappa,N\delta)$ for $\ell=0,1,2,\ldots$ is defined as follows:
\begin{subequations}
\begin{align}
\mathfrak{P}_0(\varkappa,N\delta)
&=
\sum_{k=1}^\infty \varkappa^k\mathfrak{I}_k(N\delta);
\\
\mathfrak{P}_\ell(\varkappa,N\delta)
&=
(\varkappa\partial_\varkappa)^\ell \mathfrak{P}_0
=
\sum_{k=1}^{\infty}
k^\ell \varkappa^k \mathfrak{I}_k(N\delta).
\label{eq:P_function}
\end{align}
\end{subequations}
The $I_{\text{MF}}$ for the three regimes is displayed in Fig. \ref{fig:SSB_MF_I}.
\section{Atomic Coherent States}
\subsection{Introduction}
\noindent Atomic Coherent States \cite{Radcliffe_1971,Arecchi1972atomic,Agarwal1981} are product states of identical single-atom spinors $|\theta,\phi\rangle 
= \bigotimes_{\ell=1}^N |\theta,\phi\rangle_\ell$. 
When written in the Dicke basis, they become binomially-weighted superpositions of Dicke states:
\begin{equation}
|\theta,\phi\rangle 
=\mathcal{\hat R}(\theta,\phi)|-j\rangle
=\sum_{n=0}^N C^N_n(\theta,\phi)|n-j\rangle, 
\label{ACS}
\end{equation}
where the $SU(2)$ rotation is defined as \cite{RevModPhys.62.867,Agarwal_2012}:
\begin{equation}
\mathcal{\hat R}=\exp\left\{
\frac{\theta}{2}
\Big(
e^{-i\phi}\hat J_+-e^{i\phi}\hat J_-
\Big)
\right\}
\end{equation}
and the amplitudes are given by:
\begin{equation}
C_n^N(\theta,\phi) = \sqrt{\binom{N}{n}}\cos^{N-n}(\theta/2)\sin^n(\theta/2) e^{-in\phi}.
\end{equation}

The six axial states defined on the Bloch sphere with maximal spin $|\langle \boldsymbol{\hat J}\rangle|=j$ are specific polarizations denoted by $|\pm x,y,z\rangle$. 
We adopt the convention for one atom:
\begin{subequations}
\begin{align}
&|\theta, \phi\rangle = 
\cos(\theta/2) |+z\rangle 
&&+ e^{-i\phi}\sin(\theta/2) |-z\rangle;
\\
&|+z\rangle = |0,0\rangle =|g\rangle,
&&|-z\rangle = |\pi,0\rangle = |e\rangle;
\\
&|+x\rangle = |\pi/2,0\rangle,
&&|-x\rangle = |\pi/2,\pi\rangle;
\\
&|+y\rangle = |\pi/2,\pi/2\rangle,
&&|-y\rangle = |\pi/2,3\pi/2\rangle,
\end{align}
\end{subequations}
and for $N$ atoms:
\begin{subequations}
\begin{align}
&|\pm z\rangle = |j,\mp j\rangle ; \\
&|\pm x\rangle = \frac{1}{2^j} \sum_{n=0}^{N} (\pm 1)^{n} \sqrt{\binom{N}{n}} |j,n-j\rangle; \\
&|\pm y\rangle = \frac{1}{2^j} \sum_{n=0}^{N} (\mp i)^{n} \sqrt{\binom{N}{n}} |j,n-j\rangle.
\end{align}
\end{subequations}

The general ACS is related to the RDS as follows.
Consider an arbitrarily oriented RDS $|m\rangle_\theta^\phi \equiv e^{-i\phi\hat J_z} e^{-i\theta\hat J_y} |m\rangle$ satisfying:
\begin{equation*}
\left(
\hat J_x\sin\theta\cos\phi
+
\hat J_y\sin\theta\sin\phi
+
\hat J_z\cos\theta
\right)
|m\rangle_\theta^\phi=m|m\rangle_\theta^\phi.
\end{equation*}
Then, $|\theta,\phi\rangle=e^{-ij\phi}|j\rangle_{\pi-\theta}^\phi$.

\begin{figure}
    \centering
    \includegraphics[width=1\linewidth]{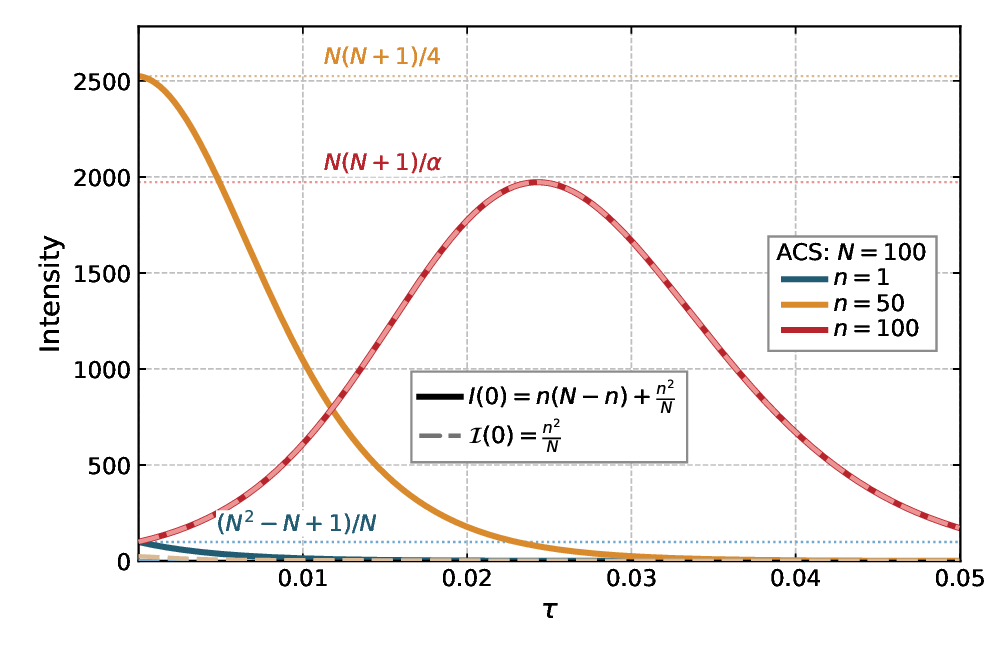}
    \caption{The total and incoherent ACS intensities.}
    \label{fig:I_ACS_all}
\end{figure}

\begin{figure}
    \centering
    \includegraphics[width=1\linewidth]{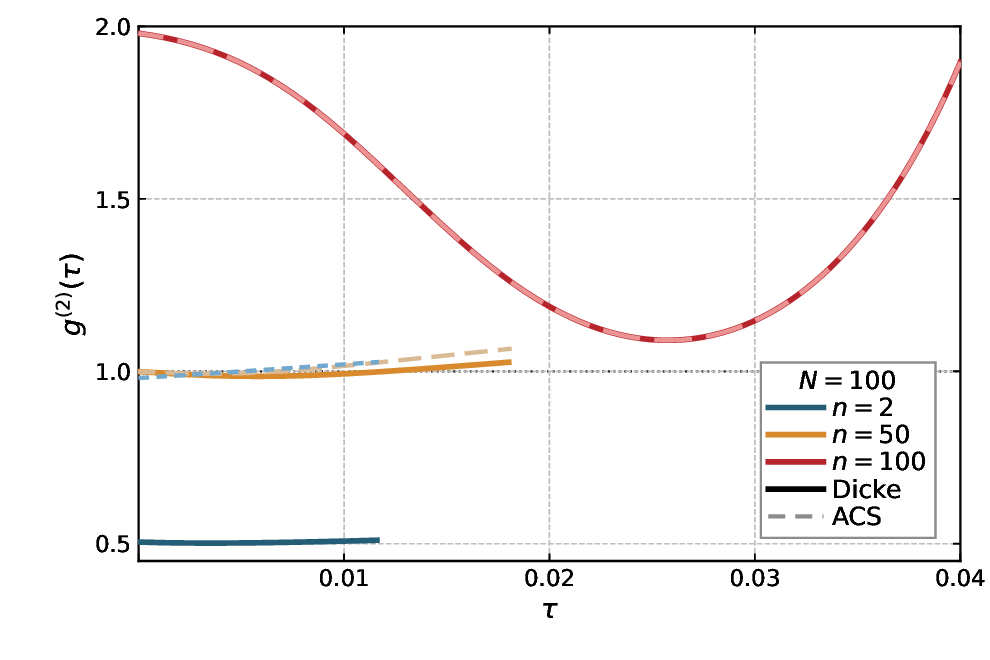}
    \caption{The Glauber function for ACS.}
    \label{fig:g_ACS_all}
\end{figure}

To find the dynamical properties of an arbitrary ACS, we write first the initial Dicke-basis populations.
Since the ACS have initial population over the full Dicke ladder including levels above $n=j$, the degeneracy in the decay rates produces repeated poles in the transfer coefficients. 
Therefore, we use \eqref{Repeated_pole_transfer} rather than the simple-pole formula used for the CDS.
The population then is given by:
\begin{equation}
P_n(0;\theta)=\binom{N}{n}\cos^{2(N-n)}(\theta/2)\sin^{2n}(\theta/2).
\end{equation}
From this follows the dynamical functions:
\begin{subequations}
\begin{align}
D^2(0;\theta)
&=
N^2\sin^2(\theta/2)\cos^2(\theta/2);
\\
I(0;\theta)
&=
N\sin^2(\theta/2)+\frac{N-1}{N}D^2(0;\theta);
\\
\mathcal I(0;\theta)
&=
N\sin^4(\theta/2);
\\
g^{(2)}(0;\theta)
&=
\frac{N-1}{N
\big[ N-(N-1)\sin^2(\theta/2) \big]^2}
\\
&\times
\big[N(N-1)-2(N-1)(N-2)\sin^2(\theta/2)
\notag \\
&+ (N-2)(N-3)\sin^4(\theta/2)\big].
\notag
\end{align}
\end{subequations}
Comparative intensity and correlation plots are given in Figs. \ref{fig:I_ACS_all} and \ref{fig:g_ACS_all}, respectively.

Choosing the $|x\rangle$ polarization, it can be immediately seen that $P_n(0)=\frac{1}{2^N}\binom{N}{n}$. 
A system initially prepared in this state has the maximal macroscopic dipole moment $D(0)=\langle \hat J_x\rangle =j$. 
Indeed, the coherent dipole contribution dominates the emission 
$\mathcal{I}(0)=I(0)-D^2(0)=\frac{N(N+1)}{4}-\frac{N^2}{4}=\frac{N}{4}$.
One notices that for large-$N$, 
\begin{equation}
g^{(2)}(0)=\frac{(N-1)(N^2+3N-2)}{N(N+1)^2}
\rightarrow 1,
\end{equation}
which is why the Poissonian limit is also called the coherent state limit.
The phase uncertainty for $|x\rangle$, is well-known, and is given by $\Delta \hat J_z(0) = \frac{\sqrt{N}}{2}$. 
Then, $\Delta \phi(0) \sim \frac{1}{\sqrt{N}}$, which is the standard quantum limit.

\subsection{Mean-Field Solution}
\begin{figure}
    \centering
    \includegraphics[width=0.96\linewidth]{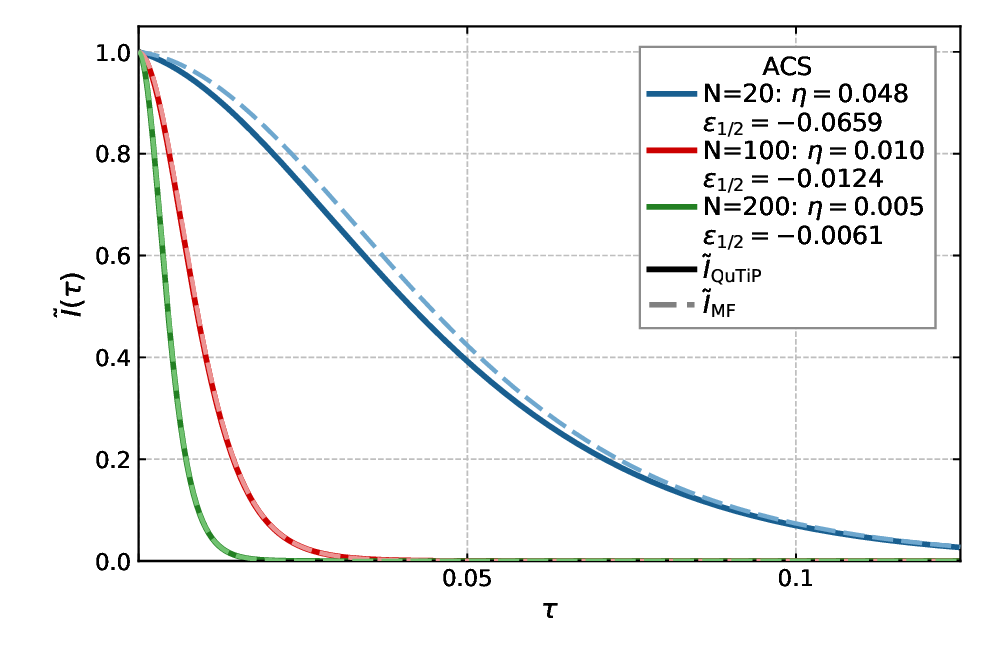}
    \caption{The ACS intensity curves. The half-width error is defined as $\varepsilon_{1/2}=(\tau_{1/2}-\tau^{\text{MF}}_{1/2})/\tau_{1/2}$.}
    \label{fig:I_ACS_Meanie}
\end{figure}

\noindent For the mean-field approximation of the $|x\rangle$ polarization, we write $n=Nx$, with $0\leq x \leq 1$, and use Stirling's approximation:
\begin{equation}
P_n(0)
\approx
\frac{2^{-N}}{\sqrt{2\pi N x(1-x)}}
\left[x^x(1-x)^{1-x}\right]^{-N}.
\end{equation}
Since the distribution is centered at $x_0=1/2$, the slowly varying square-root prefactor may be evaluated at that point. 
Expanding $f(x)=x\ln x+(1-x)\ln(1-x)$ about $x_0$, we find $f(x)=-\ln 2+\frac12(2x-1)^2$. 
Therefore, we use the de Moivre--Laplace theorem to approximate the density function:
\begin{equation*}
W(x_0,0)=
\frac{N}{2^{N}} \binom{N}{Nx_0} 
\approx 
\sqrt{\frac{2N}{\pi}}
\exp\left\{-j(2x_0-1)^2 \right\}.
\end{equation*}
Then, let us write $z=a\coth(Na\tau)+a-1$ to see that the intensity (Fig. \ref{fig:I_ACS_Meanie}) is calculated from:
\begin{align}
I_{\text{MF}}(\tau)
&\approx\sqrt{\frac{2N(N+1)^4}{\pi}}
\label{ACS_MF_integral_x0}
\\
&\times\int_0^1
\frac{x_0(a-x_0)e^{-j(2x_0-1)^2} dx_0}
{\left[a\cosh(Na\tau)+(a-2x_0)\sinh (Na\tau)\right]^2}
\notag\\
&= 
\sqrt{\frac{N(N+1)^4}{32\pi}}\csch^2(Na\tau)
\Big[
2a\coth{(Na\tau)}\mathcal F_N(z)
\notag \\
&+\left.
a^2\csch^2(Na\tau)\mathcal{F'}_N(z)
-\sqrt{2\pi/N} 
\text{erf}\left\{\sqrt{j}\right\}
\right]
\notag\\
&\xrightarrow{N\gg1}
\frac{N+1}{4}
\sech^{2}(Na\tau)
\notag\\
&\times
\Big[
N
+3\tanh^{2}(Na\tau)
-2\tanh(Na\tau)
\Big]
\notag,
\end{align}
with $N\tau_{1/2} \approx \frac{N}{N+1} \bigg[\ln(1+\sqrt2)+\frac{3\sqrt2-4}{4N}\bigg]$, 
and the error function \cite{osti_243992}, the truncated Faddeeva function \cite{poppe1990more}, and its derivative are given below:
\begin{subequations}
\begin{align}
\text{erf}\{\sqrt{j}\}
&=\sqrt{\frac{N}{2\pi}}\int_{-1}^1 dx e^{-j x^2}
\label{Error_Function}
;\\
\mathcal{F}_N(z)
&=\int_{-1}^1 \frac{dx}{z-x}e^{-j x^2}
\label{Faddeeva_integral}
;\\
\mathcal{F}'_N(z)
&=-\int_{-1}^1 \frac{dx}{(z-x)^2}e^{-j x^2}.
\label{Faddeeva_integral_Derivative}
\end{align}
\end{subequations}

\begin{comment}
$N\tau_{1/2}^{\text{MF}} = \frac{N}{N+1} \bigg[\text{arcosh}\Big(\frac{\sqrt 2 (N+1)}{\sqrt{N(N+2)}}\Big) - \ln{\sqrt{\frac{N+2}{N}}}\bigg]$
\end{comment}
\section{Discussion \& Intercomparisons}
\paragraph*{\textit{Discussion.}}
Distinct initial preparations under the same vacuum-Liouvillian can produce nearly indistinguishable total-intensity pulses, so $I(\tau)$ neither identifies the initial state nor reveals the physical mechanism generating the radiation.
At fixed $N$ and $\Gamma$, our results show that the excitation number chiefly sets the emission clock; collective rotation redistributes the initial state over the Dicke ladder and changes the coherent share of the radiation; bath squeezing controls the crossover between weakly excited and RDS-like pair-correlated emission; and ACS polarization on the Bloch sphere determines both the initial excitation and the macroscopic dipole.
Consequently, the pulse onset, peak, delay, and width, as alluded to in the introduction, must be considered together with the total and incoherent intensities as well as the Glauber function.
These observables distinguish broad classes of initiations and provide a practical guide both for the selection of an initial state that produces a desired emission profile and for diagnosing an ensemble preparation from its emitted light.

\begin{figure}
    \centering
    \includegraphics[width=1\linewidth]{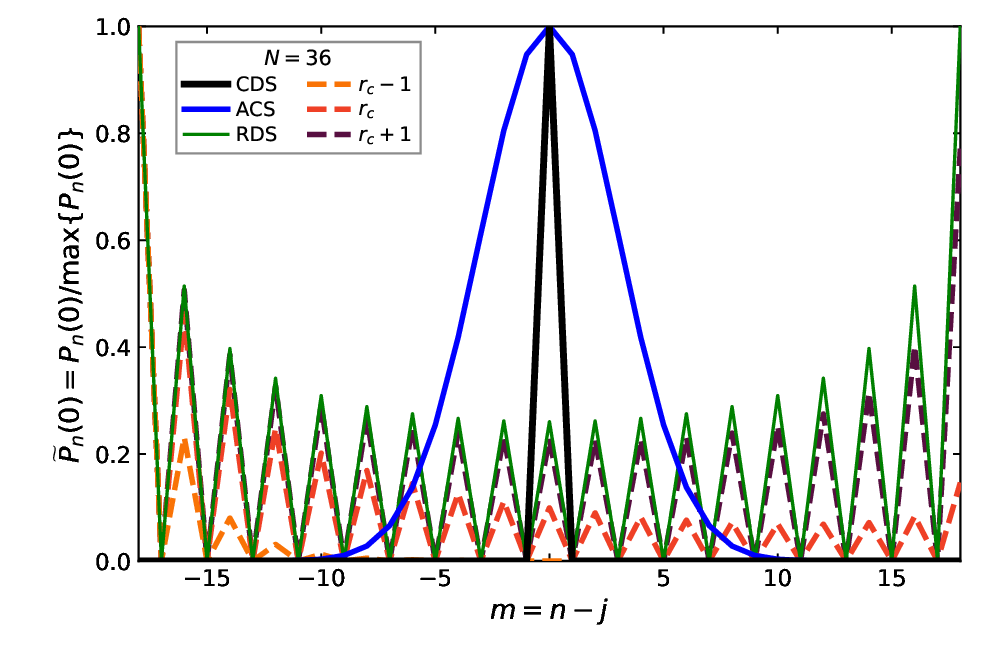}
    \caption{Normalized initial occupation probability.}
    \label{fig:pm}
\end{figure}

Fig.\ref{fig:pm} shows the initial occupation probability $P_n$ in the Dicke basis for the aforementioned initial states with $N=36$ atoms (for clarity). 
The sharply localized CDS, the broadly distributed $|x\rangle$-polarized ACS, and the RDS and SDS exhibit markedly different population profiles.
These have similar phase-insensitive dynamics and require a phase-sensitive dipole measurement (viz. $\langle \hat J_+\rangle$) to distinguish them.
The ACS further demonstrate that nearly identical pulse profiles may arise from fundamentally different emission mechanisms, since the emission from the separable ACS is predominantly carried by a macroscopic dipole, whereas the comparable entangled CDS pulse is entirely incoherent.
Note also that at large-$N$, both approach nearly the same normalized mean-field $\sech^2$ pulse, and have nearly identical fitting coefficients (table \ref{tab:largeN_main_dynamics}).
Their similarity provides a concrete demonstration that neither superradiant enhancement nor the total pulse shape constitutes an entanglement witness.
By contrast, the RDS and strongly squeezed SDS form a broad-pulse class with smaller curvature.
Decreasing the bath squeezing continuously drives the SDS toward a narrower and increasingly exponential decay.

\paragraph*{\textit{Small and Large Ensembles.}}
In the following, we present representative results that compare between the small- and large-$N$ dynamics for the different initial states discussed above. 

\begin{figure*}
    \centering
    \includegraphics[width=1\linewidth]{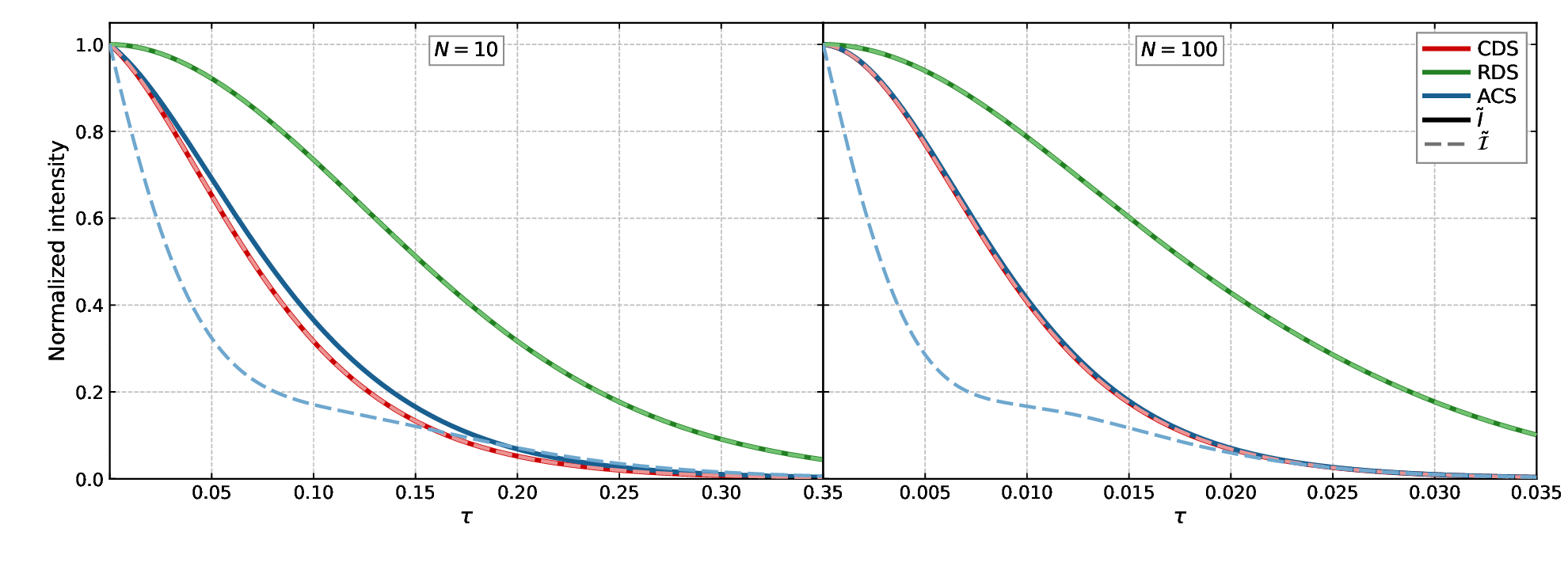}
    \caption{A comparison between the total (solid) and incoherent (dashed) intensities of CDS, central RDS, and $x$-polarized ACS for small ($N=10$) and large ($N=100$) ensembles.}
    \label{fig:All_funnyI}
\end{figure*}

\begin{figure}
    \centering
    \includegraphics[width=1\linewidth]{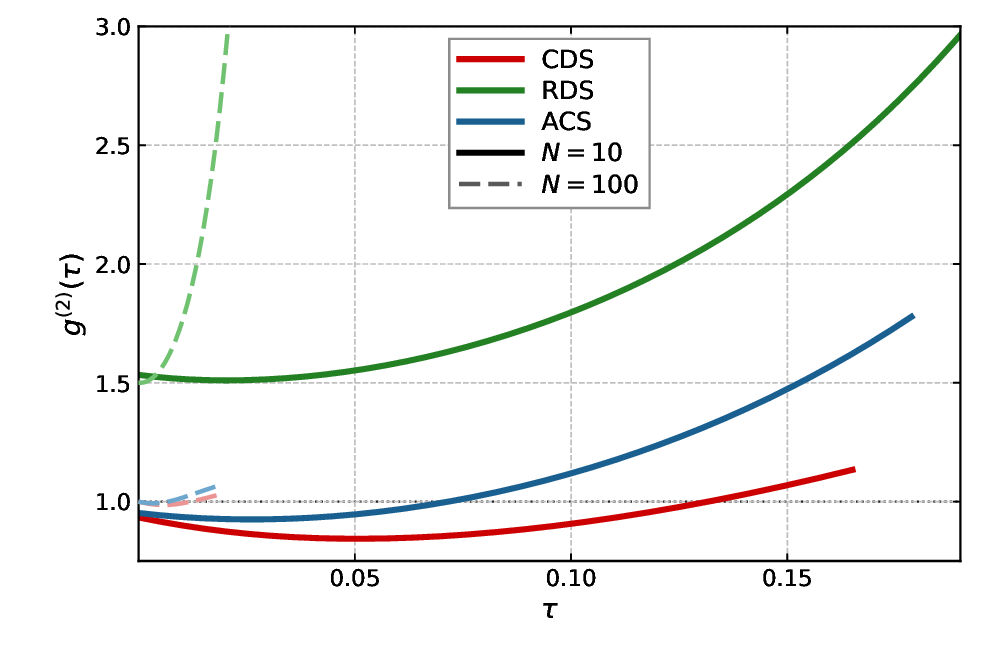}
    \caption{Comparison of the Glauber function for small ($N=10$) and large ($N=100$) ensembles.}
    \label{fig:All_g2}
\end{figure}

To guide experiments with small ensembles, Figs.~\ref{fig:All_funnyI} and \ref{fig:All_g2} compare complementary observables that distinguish the initial states even when their intensity pulses are similar.
In Fig. \ref{fig:All_funnyI}, the CDS and ACS curves have similar pulse-widths, but as noted in Sec. VI, the incoherent part of the ACS emission is initially $N+1$-times smaller than its total emission which explains the characteristic difference.
The central RDS decays more slowly because its broad initial population samples a wider range of Dicke decay channels, including slowly emitting levels near the edges of the ladder, which extend the descending tail.

As can be seen in Fig. \ref{fig:All_g2}, the central RDS is strongly bunched, whereas the CDS and ACS remain close to the Poissonian reference for large-$N$.
Our interpretation is that while both CDS and RDS \textit{store} emission strength in fluctuations, in RDS the two-photon emission intensity $\langle \hat J_+^2\hat J_-^2\rangle$ is amplified by the pair-excitation structure of the state.
Taken together, these dynamical functions provide a practical means of distinguishing the initial state in the low-$N$ regime.

\begin{figure}[htbp]
    \centering
    \includegraphics[width=1\linewidth]{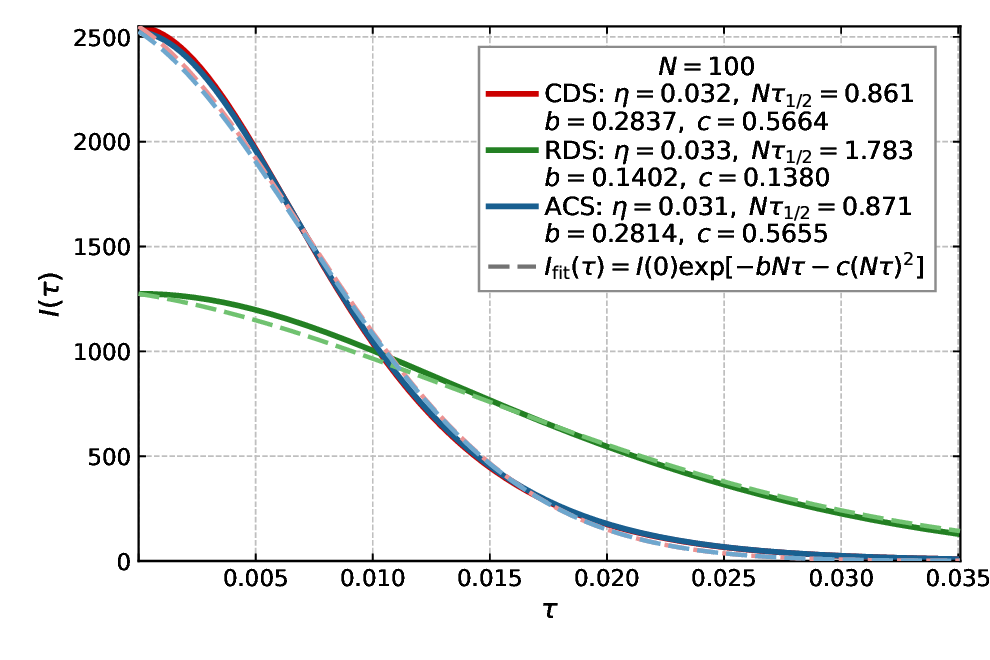}
    \caption{The intensity-fit comparison from table \ref{tab:largeN_main_dynamics}.}
    \label{fig:Gaussian_Fit}
\end{figure}

\paragraph*{\textit{Asymptotic Gaussian Fits.}}
Table \ref{tab:largeN_main_dynamics} shows the fitting parameters for the intensity.
The cutoff $\tau_{\text{cut}}$ is the last time satisfying $I(\tau_{\text{cut}}) = 0.1I_{\mathrm{peak}}$ on the final fall of the pulse.
We write $x=N\tau$ and fit the normalized intensity using both $\frac{I(\tau)}{I(0)}\simeq \exp[-\alpha x^2]$, and $\frac{I(\tau)}{I(0)}\simeq \exp[-b x-c x^2]$.
Thus, $\alpha$ is the scaled Gaussian curvature coefficient, and the pair, $b$ and $c$, are the scaled linear and quadratic coefficients of the mixed exponential--Gaussian fit.
We plot the mixed form in Fig. \ref{fig:Gaussian_Fit} for $N=500$.
This has the smaller NRMSE for all rows shown.
This is expected because the exact population solution is a finite sum of exponential (and when repeated poles occur, polynomial--exponential) decay channels rather than an exact Gaussian.
The pure-Gaussian coefficient $\alpha$ is retained because it gives a compact one-parameter measure of the pulse curvature.
For the SDS, the initial intensity at $N=500$ is given by equation \eqref{SSB_Intensity_Q}, with $\delta_{\rm strong} = -1.0\times10^{-4}$, $\delta_{\rm inter} = -8.01284\times10^{-4}$, and $\delta_{\rm weak} = -2.4\times10^{-2}$. 
\begin{table}[t]
\centering
\scriptsize
\renewcommand{\arraystretch}{1.28}
\setlength{\tabcolsep}{3.2pt}
\begin{tabular}{@{}lcccccc@{}}
\toprule
Initial state
&
$I(0)$
&
$\alpha$
&
$b$
&
$c$
&
$N\tau_{1/2}$
&
$N\tau_{\text{cut}}$
\\
\midrule

$|j,0\rangle$
&
$N(N+2)/4$
&
$0.7617$
&
$0.2837$
&
$0.5664$
&
$0.8772$
&
$1.8151$
\\

$|j,0\rangle_x$
&
$N(N+2)/8$
&
$0.1859$
&
$0.1402$
&
$0.1380$
&
$1.7714$
&
$3.6587$
\\

$|j,0\rangle_x^{\delta_{\rm strong}}$
&
$I(0;\delta_{\rm strong})$
&
$0.1922$
&
$0.1646$
&
$0.1353$
&
$1.7205$
&
$3.6123$
\\

$|j,0\rangle_x^{\delta_{\rm inter}}$
&
$I(0;\delta_{\rm inter})$
&
$0.2471$
&
$0.3374$
&
$0.1179$
&
$1.3891$
&
$3.2632$
\\

$|j,0\rangle_x^{\delta_{\rm weak}}$
&
$I(0;\delta_{\rm weak})$
&
$1.9411$
&
$1.7930$
&
$0.0921$
&
$0.3803$
&
$1.2116$
\\

$|x\rangle$
&
$N(N+1)/4$
&
$0.7589$
&
$0.2814$
&
$0.5655$
&
$0.8794$
&
$1.8179$
\\

\bottomrule
\end{tabular}
\caption{
Large-$N$ fitting summary at $N=500$. 
}
\label{tab:largeN_main_dynamics}
\end{table}

\section{Conclusion}
We have shown in this article that the initial state is a decisive control parameter in the subsequent superradiant dynamics.
The coherent and incoherent contributions to the emission can be analyzed to differentiate between states with similar total intensity profiles.
In addition, the quantum statistical characteristics (via the Glauber correlation function) are studied to highlight the differences between different initial states.
A method of simplifying the superradiant master equation analytically and approximately (through the Fokker--Planck equation) was developed for all the initial states considered here. 
The differences in emission and correlation properties between small and large ensembles have been considered, as small ensembles are relevant for studies with superconducting qubits, and large ensembles to atomic vapors.

\section*{Data Availability Statement}
\noindent Data used herein were generated by methods described within. Codes will be provided upon reasonable request.

\begin{acknowledgments}
\noindent All numerical fits were done through \verb|SciPy|, \verb|NumPy|, and \verb|QuTiP| \cite{2020SciPy-NMeth, harris2020array, qutip5}. AA would like to thank Dr. Abdullah Al-Shammari from Kuwait University for many helpful suggestions. ZZ is supported by the Heep Graduate Fellowship. We are grateful for the support from the Robert A. Welch Foundation (Grant No. A-1943-20240404).
\end{acknowledgments}

\section*{Author Contributions}
All authors contributed equally to this work.

\appendix

\section{Solving the Master Equation}
\noindent
Here, we provide the general method by which we have solved our master equations. The superradiance master equation \cite{Agarwal_Master_Equation} is given by:
\begin{equation}
\partial_t \rho 
= \mathcal{L}\{\rho\}
= -\Gamma \left(
\hat J_+ \hat J_- \rho
- 2\hat J_- \rho \hat J_+ 
+ \rho \hat J_+ \hat J_-
\right),
\label{Master_Equation}
\end{equation}
where $\mathcal{L}\{\bm{\cdot}\}$ is the Liouvillian operator. It takes as its argument the density matrix and evolves it in time. We take the initial state, $|\psi_0\rangle$, and construct from it the density matrix $\rho(0)=|\psi_0\rangle \langle \psi_0|$. Since Dicke states constitute a complete basis for symmetric spin states, we expand our density matrices in terms of coefficients $\rho_{nn'}$ in the Dicke basis such that:
\begin{align}
\rho(t) 
&=
\sum_{n, n'=0}^N
\rho_{n n'}(t)
|j,n-j\rangle \langle j, n'-j|\notag \\
&=
\sum_{n=0}^{N} \rho_{nn}(t) |j, n-j\rangle \langle j, n-j|\notag \\
&+
\sum_{n \neq n'} \rho_{nn'}(t) |j, n-j\rangle \langle j, n'-j|.
\label{Density_Matrix_Dicke_t}
\end{align}
Note that we have used the excitation number representation $n = m+j$. The first series on the RHS of \eqref{Density_Matrix_Dicke_t} is the population matrix, with $\rho_{nn}(t)=P_n(t)$ and $\sum_{n=0}^NP_n(t)=1$, whereas the off-diagonal elements, $\rho_{nn'}(t)$, are the coherences, representing the quantum superpositions between states. 

Then, we can plug \eqref{Density_Matrix_Dicke_t} into \eqref{Master_Equation},with $|n\rangle \equiv |j,n-j\rangle$, and sum from $n=0\rightarrow N$. The population equation satisfies:
\begin{align}
\dot P_n(t)|n\rangle \langle n|
&=
2\Gamma P_n(t) \left( \hat J_- |n\rangle \langle n| \hat J_+ - \frac{1}{2} \left\{\hat J_+ \hat J_-, |n\rangle \langle n| \right\} \right)
\notag \\
&=
\lambda_n P_n(t) \left( |n-1\rangle \langle n-1|  - |n\rangle \langle n| \right),
\label{Dicke_Master_1}
\end{align}
with $\lambda_n = 2\Gamma n(N-n+1)$ being the decay rates. If we shift the series on the RHS by 1, and then trace over these states, we get the indicial rate equation:
\begin{equation}
\dot P_n(t) =  \lambda_{n+1} P_{n+1}(t)-\lambda_{n} P_{n}(t),
\label{Indicial_Dicke}
\end{equation}
which may be solved via the Laplace transform method, where $\tilde P_n(s)=\mathfrak L\{P_n(t)\}(s)=\int_0^\infty e^{-st} P_n(t)dt$, $\mathfrak L\{\dot P_n(t)\}=s\tilde P_n(s)-P_n(0)$, and where as usual, $P_n(0)$ is the initial probability that the system starts in level $n$. Transforming \eqref{Indicial_Dicke}, we get after rearranging:
\begin{equation}
\tilde P_n(s) = \frac{P_n(0)+\lambda_{n+1} \tilde P_{n+1}(s) }{s+\lambda_n}.
\label{Laplace_Indicial}
\end{equation}
This is our fundamental recursion relation. If we carry out this recursion until we arrive at $n=0$, we get:
\begin{equation}
\tilde P_n(s)=
\sum_{k=n}^N P_k(0)
\overbrace{\left[
\frac{\prod_{r=n+1}^k \lambda_r}{\prod_{\ell=n}^k(s+\lambda_\ell)}
\right]}^{=\tilde C_{n,k}(s)},
\label{Laplace_Rates}
\end{equation}
where $\tilde C_{n,k}(s)$ are the transfer coefficients, whose inverse-Laplace transform is the probability that a system starting in state $k$ has decayed into state $n$ by time $t$. Now, a subtle point: Due to the degeneracy of the decay rates $\lambda_n=\lambda_{N-n+1}$, we may get a pole of order 1 or 2. For poles of order 1, as we find in our Dicke states, we find that the transfer coefficients are:
\begin{equation}
C_{n,k}(t) = 
\left( \prod_{r=n+1}^{k} \lambda_r \right) \sum_{\ell=n}^{k} \frac{e^{-\lambda_\ell t}}{\prod_{\substack{q=n \\ q \neq \ell}}^{k} (\lambda_q - \lambda_\ell)}.
\label{Dicke_Transfer}
\end{equation}

To address states with pole of order q, we note first that:
\begin{equation*}
\mathfrak{L}^{-1}\left\{
\frac{1}{(s+\lambda_\ell)^q}
\right\} (t)
=
\frac{t^{q-1}}{(q-1)!}e^{-\lambda_\ell t}.
\end{equation*}
We define the set $\mathcal{M}_{n,k}$ of distinct rates and the multiplicity $m_{n,k}$ of the distinct rates:
\begin{subequations}
\begin{align}
&\mathcal{M}_{n,k}=\{\lambda_\ell: \quad \ell = n, n+1, \dots, k-1, k\}; 
\\ 
&m_{n,k}(\lambda_\ell \in \mathcal{M}_{n,k})=\sum_{\ell'=n}^k \delta_{\lambda_{\ell'},\lambda_\ell };
\label{multiplicity_A}
\\
&
\mathcal M_{n,k}^{(q)}
=
\{\lambda_\ell\in\mathcal M_{n,k}: m_{n,k}(\lambda_\ell)=q\}.
\end{align}
\end{subequations}
Since no state considered in this article has a pole higher than the second order, we may write:
\begin{align}
C_{n,k}(t)
&=
\sum_{\mu\in\mathcal M_{n,k}^{(1)}}A^{(1)}_{n,k}(\mu)e^{-\mu t}
\notag
\\
&+
\sum_{\mu\in\mathcal M_{n,k}^{(2)}}
\bigl(A^{(2)}_{n,k}(\mu)t + A^{(3)}_{n,k}(\mu)\bigr)e^{-\mu t},
\label{Repeated_pole_transfer}    
\end{align}

where the weights $A^{(i)}_{n,k}(\mu)$ are given by:
\begin{subequations}
\begin{align}
A^{(1)}_{n,k}(\mu\in\mathcal M_{n,k}^{(1)})
&=
\left[
(s+\mu)\,\tilde C_{n,k}(s)
\right]_{s=-\mu};
\\[0.75em]
 A^{(2)}_{n,k}(\mu\in\mathcal M_{n,k}^{(2)})
&=
\left[
(s+\mu)^2\,\tilde C_{n,k}(s)
\right]_{s=-\mu};
\\[0.75em]
 A^{(3)}_{n,k}(\mu\in\mathcal M_{n,k}^{(2)})
&=
\left[
\partial_s\left((s+\mu)^2\,\tilde C_{n,k}(s)\right)
\right]_{s=-\mu}.
\end{align}
\end{subequations}

Then, the populations can be read from \eqref{Laplace_Rates} to be:
\begin{equation}
P_n(t)=\sum_{k=n}^N P_k(0) C_{n,k}(t).
\label{Populations_2}
\end{equation}

Now, for the coherences. We may write for $\Delta \in [0, N]$:
\begin{equation}
\rho^{(\Delta)}_n(t)\equiv\rho
_{n,n+\Delta}(t), 
\end{equation}
with the allowed transitions being between $0\leq n\leq N-\Delta$. We notice immediately that $\Delta = 0$ was the population band just addressed. (We find that it is more transparent to keep these methods separate.) In general, $\Delta = \ell$ gives us the moments $\langle \hat J_+^\ell\rangle$. Then, the coherences separately solve:
\begin{equation}
\dot \rho^{(\Delta)}_n(t) 
=
\alpha_n^{(\Delta)} \rho_{n+1}^{(\Delta)}(t)-\beta_n^{(\Delta)}\rho_n^{(\Delta)}(t),
\end{equation}
with $\alpha_n^{(\Delta)}=\sqrt{\lambda_{n+1}\lambda_{n+\Delta+1}}$ and $\beta_n^{(\Delta)}=\frac{\lambda_n+\lambda_{n+\Delta}}{2}$. Following the same method, we take the Laplace transform and rearrange to get:
\begin{equation}
\tilde{\rho}_n^{(\Delta)}(s)
=
\sum_{k=n}^{N-\Delta}
\rho_k^{(\Delta)}(0)\,
\frac{\displaystyle\prod_{r=n}^{k-1}\alpha_r^{(\Delta)}}
{\displaystyle\prod_{\ell=n}^{k}\bigl(s+\beta_\ell^{(\Delta)}\bigr)},
\end{equation}
and then take the inverse Laplace transforms:
\begin{equation}
\mathfrak{L}^{-1}\!\left\{
\overbrace{\frac{1}{\bigl(s+\beta_{\ell}^{(\Delta)}\bigr)^q}}^{\tilde C_{n,k}^{(\Delta)}(s)}
\right\}(t)
=
\frac{t^{q-1}}{(q-1)!}\,e^{-\beta_{\ell}^{(\Delta)} t},
\end{equation}
to find that the transfer coefficients are:
\begin{align}
C_{n,k}^{(\Delta)}(t)
&=
\sum_{\beta_\ell^{(\Delta)}\in\mathcal M_{n,k}^{(1,\Delta)}}
A_{n,k}^{(1,\Delta)}\!\bigl(\beta_\ell^{(\Delta)}\bigr)e^{-\beta_\ell^{(\Delta)} t}
\notag \\
&+
\sum_{\beta_\ell^{(\Delta)}\in\mathcal M_{n,k}^{(2,\Delta)}}
\left[
A_{n,k}^{(2,\Delta)}\!\bigl(\beta_\ell^{(\Delta)}\bigr)t
\right.
\notag\\
&+
\left.
A_{n,k}^{(3,\Delta)}\!\bigl(\beta_\ell^{(\Delta)}\bigr)
\right]
e^{-\beta_\ell^{(\Delta)} t}.
\end{align}

Finally, the coherences will be:
\begin{equation}
\rho_n^{(\Delta)}(t)
=
\sum_{k=n}^{N-\Delta}
\rho_k^{(\Delta)}(0)\,
C_{n,k}^{(\Delta)}(t).
\label{Coherences}
\end{equation}
Then, since: 
\begin{align*}
\hat J_+ |j, m\rangle 
&= \sqrt{(j-m)(j+m+1)}|j,m+1\rangle
\\
&=\sqrt{\frac{\lambda_{n+1}}{2\Gamma}}|j, n-j+1\rangle
;\\
\hat J_+ \hat J_-|j,m\rangle 
&=(j+m)(j-m+1)|j,m\rangle
=\frac{\lambda_n}{2\Gamma}|j,n-j\rangle
;\\
\hat J_+^2J_-^2 |j,m\rangle 
&= (j+m)(j^2-(m-1)^2)(j-m+2)|j, m\rangle 
\\
&= \frac{\lambda_n \lambda_{n-1}}{(2\Gamma)^2}|j, n-j\rangle
.
\end{align*}
we can find the expressions of the dipole moment $D(\tau)=|\langle\hat J_+\rangle|(\tau)$, the total emission intensity $I(\tau)=\langle\hat J_+\hat J_-\rangle(\tau)$, and the Glauber function $g^{(2)}(\tau)=\frac{\langle\hat J_+^2\hat J_-^2\rangle(\tau)}{\langle\hat J_+\hat J_-\rangle^2(\tau)}$ from:
\begin{align}
&\label{Dipole_Moment_analytical}
D(\tau)
=
\left|
\sum_{n=0}^{N-1}\sqrt{\frac{\lambda_{n+1}}{2\Gamma}}\,\rho_{n,n+1}(\tau)
\right|
;\\
&\label{Total_Intensity_analytical}
I(\tau)
=
\sum_{n=0}^{N}\frac{\lambda_n}{2\Gamma}\,P_n(\tau)
;\\
&\label{Glauber_function_Analytical}
g^{(2)}(\tau)
=
\frac{
\displaystyle\sum_{n=2}^{N}
\lambda_n\lambda_{n-1}
P_n(\tau)
}{
\left(
\displaystyle\sum_{n=0}^{N}
\lambda_n
P_n(\tau)
\right)^2
}
.
\end{align}

\begin{comment}
Finally, the phase uncertainty is calculated from:
\begin{equation}
\Delta \phi \sim \frac{1}{2\Delta J_z} = \frac{1}{2\sqrt{\langle \hat J_z^2\rangle - \langle \hat J_z\rangle^2} },
\end{equation}
where:
\begin{equation}
\langle \hat J_z^\ell\rangle (\tau)
=
\sum_{n=0}^N (n-j)^\ell P_n(\tau).
\end{equation}
\end{comment}
\section{Solving the Fokker-Planck Equation}
\noindent We may give approximate solutions that are accurate for large values of $N$ by reducing our Rates equation to a Fokker-Planck (FP) equation \cite{risken1996fokker}. These solutions are often called Mean-Field Approximations, when we neglect the diffusion and take the large $N$-limit. 

Let $x=\frac{n}{N}$ be the excitation fraction. For large $N$, $x$ may be treated as a continuous variable on $[0/N,N/N]=[0,1]$. Then, we can see that 
$ \Delta x = \frac{n+1}{N} - \frac{n}{N} = \frac 1 N $. If we define $a=1+1/N$, then the decay rates are found to be:
\begin{equation}
\lambda_n = 2\Gamma N^2 \frac{n}{N} \big( 1 - \frac n N + \frac{1}{N} \big) = 2\Gamma N^2 x(a-x)\equiv \lambda(x).
\label{Lambda_x}
\end{equation}

Let us now define a continuum (mean-field) density function $W(x,\tau)$ such that $\int_0^1 dx W(x,\tau)=1$. To associate this density with our discrete $n$ population rates, we have to write the density as a piecewise discrete function as a delta-comb measure to get the pair:
\begin{subequations}
\begin{align}
&W(x,\tau)=\sum_{n=0}^N P_n(\tau) \delta\left( x-\frac{n}{N} \right);
\\
&P_n(\tau)=\int_{x-\frac{\Delta x}{2}}^{x+\frac{\Delta x}{2}} dx W(x,\tau). 
\end{align}
\end{subequations}
We can now see that:
\begin{equation}
NP_n(\tau)\approx W(x,\tau)N\Delta x = W(x,\tau),
\end{equation}
Now, a standard result from Operator Calculus gives us for a constant $\alpha$, the shift operator $e^{\alpha \partial_x}f(x)$, or the generator of infinitesimal translations:
\begin{align}
e^{\alpha\partial_x}f(x) 
&= 
\left(
1+\alpha\partial_x + \frac{\alpha^2}{2!}\partial^2_x+\dots
\right)f(x)
\notag \\
&=
\left(
1+\frac{1}{1!}(x+\alpha-x)^1\partial^1_x +\dots
\right)f(x+\alpha-\alpha)
\notag \\
&=f(x+\alpha),
\end{align}
where we notice from the last step that this is the Taylor expansion of $f(x+\alpha)$ around $x$. Now, let us take the time derivative of the continuum distribution function:
\begin{equation*}
\partial_\tau W(x,\tau)=N \dot P_n(\tau) = N[\lambda_{n+1}P_{n+1}(\tau)-\lambda_nP_n(\tau)],
\end{equation*}
where we used the rate equation. Then, we check that:
\begin{equation}
\lambda_{n+1}
= 2\Gamma N^2 [(x+\Delta x)(a-(x+\Delta x))] = \lambda(x+\Delta x).
\end{equation}

Whereas $P_{n+1}(\tau) = \frac{W(x+\frac{1}{N}, \tau)}{N}=\Delta xW(x+\Delta x, \tau)
$. 
Putting it all together, we have:
\begin{equation*}
\frac{\partial_\tau W(x,\tau)}{N\Delta x} = \left[
\lambda(x+\Delta x)W(x+\Delta x, \tau) - \lambda(x)W(x,\tau)
\right].
\end{equation*}

If we let $F(x,\tau)=\lambda(x)W(x,\tau)$, then we see that $F(x+\Delta x,\tau)=e^{\Delta x \partial_x}F(x,\tau)$. 
Then:
\begin{align}
\partial_\tau W(x,\tau) 
&= 
F(x+\Delta x, \tau) - F(x,\tau)
=\left(e^{\Delta x\partial_x} - 1\right) F(x,\tau)
\notag \\
&=
\frac{\Delta x}{1!} \partial_x F(x,\tau)
+ \frac{\Delta x^2}{2!} \partial^2_x F(x,\tau) + \dots
\end{align}
This reduces to what is known as the Kramers--Moyal Series, with Kramers--Moyal coefficients, respectively written as:
\begin{subequations}
\begin{align}
\partial_\tau W(x,\tau) 
&= \sum_{k=1}^\infty (-\partial_x)^k 
\left[
D^{(k)}(x,\tau) W(x,\tau)
\right];\\
D^{(k)}(x,\tau)
&= (-1)^k\frac{(\Delta x)^k}{k!}\lambda(x).
\end{align}
\end{subequations}

The FP equation ignores all terms $k$ higher than 2. Then, the drift $d(x,\tau)$ and diffusion $D(x,\tau)$ coefficients are given by:
\begin{subequations}
\begin{align}
&d(x,\tau) = D^{(1)}(x,\tau) = - \frac{\lambda(x)}{\Gamma N}=2 N x(x-a);\\
& D(x,\tau)=D^{(2)}(x,\tau)=\frac{\Delta x^2}{2\Gamma N^2}\lambda(x) = x(a-x).
\end{align}
\end{subequations}

The FP equation can be written as:
\begin{align}
\partial_\tau W(x,\tau)
&= -\partial_x[d(x,\tau)W(x,\tau)] + \partial_x^2 [D(x,\tau)W(x,\tau)]
\notag \\
&=
2 N
\partial_x [x(a-x)W]+\partial_x^2[x(a-x)W]
\notag
\end{align}

Now, we may set the boundary conditions. Whereas $W(x,0)$ is state-dependent, the boundary conditions depend on processes within the context of the FP equation. First, let us consider the former:
\begin{equation}
W(x,0)=\sum_{n=0}^N P_{N x'}(0) \delta(x-x'),
\end{equation}
where $N x' = n'$. Now, since $I(\tau)=\sum_{n=0}^N \frac{\lambda_n}{2\Gamma} P_n(\tau)$, we can extend this to:
\begin{equation}
I(\tau) \approx \frac{1}{2\Gamma}\int_0^1 dx \lambda(x) W(x,\tau).
\label{FP_Mean_Field_Intensity}
\end{equation}

Let us define the probability-current:
\begin{equation}
J(x,\tau)= d(x,\tau) W(x,\tau)- \partial_x [D(x,\tau)W(x,\tau)],
\end{equation}
where we set as our boundary condition the conservation of probability current $J(0,\tau)=J(1,\tau)=0$, then solve $\partial_\tau W(x,\tau)=-\partial_x J(x,\tau)$. This preserves $\int_0^1 dx W(x,\tau)=1$, which allows us to shift the integration to the initial time through $W(x_0,0)=W(x(\tau),\tau)\Big|\frac{dx}{dx_0}\Big|$.

%\nocite{*}

\bibliography{references}% Produces the bibliography via BibTeX.

\end{document}